\documentclass[%
 reprint,
superscriptaddress,
 amsmath,amssymb,
 aps,
]{revtex4-2}
\usepackage{xcolor}
\usepackage{graphicx}
\usepackage{dcolumn}
\usepackage{bm}
\usepackage{bbm}
\usepackage{relsize}
\usepackage{romannum}
\usepackage{csquotes}
\usepackage{xcolor}
\usepackage{soul}
\usepackage{floatrow}
\usepackage[mathscr]{eucal}
\usepackage{physics}

\begin{document}

\preprint{APS/123-QED}

\title{Quantum Control of Heat Current}
\author{Gobinda Chakraborty}
 \affiliation{Centre for Quantum Engineering, Research and Education, TCG CREST, Salt Lake, Kolkata 700091, India}
\author{Subhadeep Chakraborty}%
 \email{c.subhadeep91@gmail.com}
\affiliation{Centre for Quantum Engineering, Research and Education, TCG CREST, Salt Lake, Kolkata 700091, India}

\author{Tanmoy Basu}
\affiliation{Centre for Quantum Engineering, Research and Education, TCG CREST, Salt Lake, Kolkata 700091, India}
\author{Manas Mukherjee}
 \email{cqtmukhe@nus.edu.sg}
\affiliation{Centre for Quantum Technologies, National University Singapore, Singapore, 117543, Singapore}
\affiliation{NQFF, Institute of Material Research and Engineering, Agency for Science, Technology and Research, 2 Fusionopolis Way, Singapore 138634}


\begin{abstract}
Controlled flow of heat at the microscopic or nanoscopic level requires a deeper understanding of thermodynamics at the quantum scale. In a modelled trimer system formed by three quantum harmonic oscillators, we show that the local heat flow between them can be fully controlled by a single degree of freedom. The unique use of the quantum phase of the coupling interaction, allows the control to be both precise and robust. Our investigation reveals the possibility of an atypical controlled heat flow (both in direction and magnitude) between two bodies in thermal equilibrium without violating the zeroth law. Our analysis is further carried over the strong coupling regime where we report a novel phase reversal of the nonequilibrium current. The model quantum system therefore opens up new possibilities to implement quantum-controlled heat devices which we refer to as {\it phonon electronics}.
\end{abstract}

\maketitle

{\it Introduction:} The field of quantum thermodynamics~(QTD) \cite{vinjanampathy2016quantum, kosloff2013quantum} has emerged as a promising area of research that aims to extend our understanding of the thermodynamic processes into the realm of quantum mechanics, with an aim of translating to new technologies. While classical thermodynamics has been remarkably successful in describing macroscopic phenomena and using it for technology implementation, it encounters limitations when applied to systems at the microscopic level. This opens up new avenues for investigating quantum heat engines (refrigerators)~\cite{bouton2021quantum, levy2012quantum}, quantum-controlled heat flow \cite{ronzani2018tunable, senior2020heat, mandarino2021thermal, saaskilahti2013thermal}, and work extraction in the quantum realm~\cite{vsafranek2023work}. 

As a starting point, the study of heat transport in quantum systems generically adopts an open quantum system description where the quantum system under study is linked to two or more baths. Such a system is frequently described by the quantum master equation (QME) where each bath induces a jump operator to their respective sites The situation gets even more complex when the system comprises several interconnected subsystems, each coupled to separate baths. However, Ref~\cite{levy2014local} reports that the mere adoption of {\it local} QME could lead to thermodynamic inconsistencies Thus while dealing with multi-party quantum systems, a {\it global} approach naturally arises in which the master equation is derived by considering full-system Hamiltonian. Notably, in recent times such a {\it local vs. global} QME have gained considerable interest, validating their thermodynamic consistencies~\cite{rivas2010markovian, cattaneo2019local, de2018reconciliation}.

In parallel, the role of geometry in shaping the behaviour of transport in quantum systems has been studied in ~\cite{lai2018tunable, dugar2022geometry}. Notably, the triangular geometry is the minimal model that exhibits \enquote{multiple paths}, which is necessary for quantum interference effects. This triangular loop may also enclose a non-trivial phase to show Aharonov-Bohm-like effect~\cite{aharonov1959significance}. This can be achieved with the help of a synthetic magnetic field in different platforms, including transmon superconducting qubit~\cite{roushan2017chiral}, photonic lattices \cite{fang2012realizing} ultracold neutral atoms \cite{lin2009synthetic} and trapped ion \cite{kiefer2019floquet}. This study also demonstrates the inhibiting effect of dark states on quantum transport due to their ‘non-absorbing’ nature that arises from the destructive quantum interference~\cite{scully1992high}. The suppression of transport properties due to these states can be lifted by higher-order tunnelling processes~\cite{weymann2011dark} and synthetic gauge fields~\cite{emary2007dark} leading to non-reciprocal transport in the systems which has been employed in a wide range of nanophotonic devices including optical isolators~\cite{zhang2018thermal}, circulators~\cite{scheucher2016quantum}. The nonequilibrium properties of quantum trimer with a gauge field are mostly explored in the weak coupling regime where the quantum baths act locally on their respective sites.  Thus it will be quite intriguing to ask what happens in the strong coupling regime where the internal couplings dominates over the system-bath coupling.


In this letter, we explore quantum interference in an open bosonic trimmer. We start with a weakly coupled system-bath model where the master equation describes the dynamics of the reduced system. Our study reports an atypical nonequilibrium nonreciprocal current between two initially thermally equilibrated bodies. We then go beyond the weak coupling regime and report a novel phase reversal in the nonequilibrium nonreciprocal current.

\begin{figure}[!t]
   \begin{center}
   \includegraphics[ width = 8 cm]{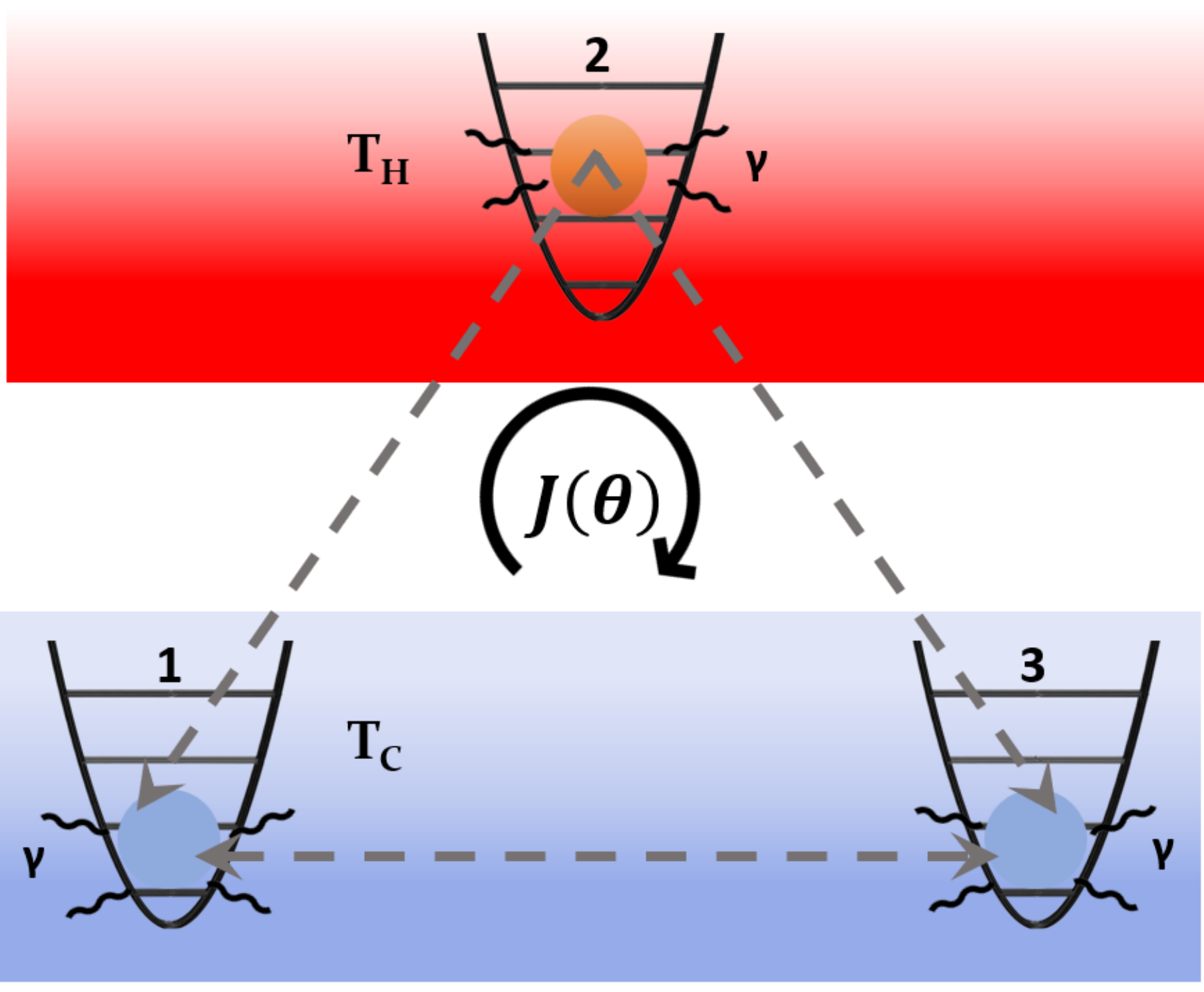}
   \caption{Schematic illustration of the system: Three quantum harmonic oscillators (labelled 1, 2, and 3) are interconnected in a loop via a phase ($\theta$)-dependent exchange interaction $J$ and are coupled to thermal baths through system-bath coupling strengths denoted by $\gamma$. Oscillators 1 and 3 are in contact with a thermal bath at temperature $T_C$, while oscillator 2 is coupled to a thermal bath at a higher temperature $T_H$  $(T_H >T_C)$. The dashed lines denote the flow of current between 1 and 3 which can be bi-directional depending on the value of $\theta$.}  
   \label{fig:1}    
   \end{center}
  \end{figure}

{\it System and its nonequilibrium steady-state:}
The system we analyze is shown in Fig.~\ref{fig:1} where three quantum harmonic oscillators (QHO), arranged in a triangular configuration, are coupled through a phase-dependent exchange interaction. The Hamiltonian of such a system can be described by
\begin{equation}\label{Ham}
        \emph{H} = \sum_{l=1} ^3 \omega a_l ^\dagger a_l +  \sum_{lm} J_{lm} \left(e^{i \theta_{lm}} a_l ^ \dagger a_m + e^{-i \theta_{lm}} a_l a_m ^\dagger \right),
\end{equation}
where $a_l$ and $a^\dagger_l$ are the annihilation and creation operators of the $l$-\textit{th} QHO (with frequency $\omega$), while $J_{lm}$ and $\theta_{lm}$ are respectively the amplitude and phase, at which a boson hops between the site $l$ and $m$. For simplicity, we assume an equal hopping amplitude (unless mentioned otherwise) $J_{lm}=J$ and introduce a cumulative phase parameter $\theta=\theta_{12} + \theta_{23} + \theta_{31}$. The Hamiltonian \eqref{Ham} yields a twofold degenerate eigenspectrum at $\theta=n\pi$ ($n$ being an integer), lying in the range $\omega-2J\leq \omega_n\leq \omega+2J$. Changing $\theta$ to any non-integer multiple of $\pi$ lifts the degeneracy, through an asymmetric exchange between sites $1$ and $3$.

In this letter, we are motivated to study the nonequilibrium aspects of such a system. To this end,  we bring the entire system in contact with two heat reservoirs, at different temperatures. The driven-dissipative dynamics~\cite{breuer2002theory} of the system is given by the following Lindblad quantum master equation
\begin{equation}\label{rho_dot}
     \frac{d\rho}{dt} = -i [\emph{H},\rho] + \sum_{k=1}^{3} \mathcal{D}[L_k](\rho),
\end{equation}
where $\mathcal{D}[L_l](\rho) = \gamma_{l} \left(N_{l}+1\right)\left(L_l\rho L_{l}^{\dagger}-\frac{1}{2}\{L_{l}^{\dagger}L_{l},\rho\}\right) + \gamma_{l} N_{l}\left(L_{l}^{\dagger}\rho L_{l} - \frac{1}{2}\{L_{l} L_{l}^{\dagger},\rho\})\right)$ represent the Lindblad superoperator with $L_l=\{a_l\}$, $N_l = \left[\textrm{exp}\left(\omega/T_l\right)-1\right]^{-1}$ denote the mean number of bosons at temperature $T_l\in\{T_C,T_H\}$, and $\gamma_l$ account for the system-bath coupling rates. Under a temperature bias $T_H>T_C$, the system reaches a nonequilibrium steady-state $\lim_{t\to \infty}\rho(t)=\rho^{NESS}$ where local thermal current flows across the system. From the continuity equation, we find this local current operator ($l\rightarrow m$) to be~\cite{landi2022nonequilibrium}
\begin{equation}\label{current}
\mathcal{J}_{lm} = i J_{lm} \left(e^{i \theta_{lm}}a^\dagger_{l}a_m - e^{-i \theta_{lm}}a_m ^\dagger a_l\right),
\end{equation}
whose expectation value are obtained as $\langle \mathcal{J}_{lm} \rangle = Tr\left(\rho^{NESS} \mathcal{J}_{lm}\right)$. The expectation of the current operator can then be obtained as $\langle \mathcal{J}_{lm} \rangle = Tr\left(\rho^{NESS} \mathcal{J}_{lm}\right)$. We note that these non-equilibrium currents are fundamentally different from the equilibrium chiral currents~\cite{roushan2017chiral}, which involves the ground state expectation of $\langle \mathcal{J}_{lm}\rangle$. 

\begin{figure}[!t]
\begin{center}
\includegraphics[ height = 6 cm, width = 6 cm]{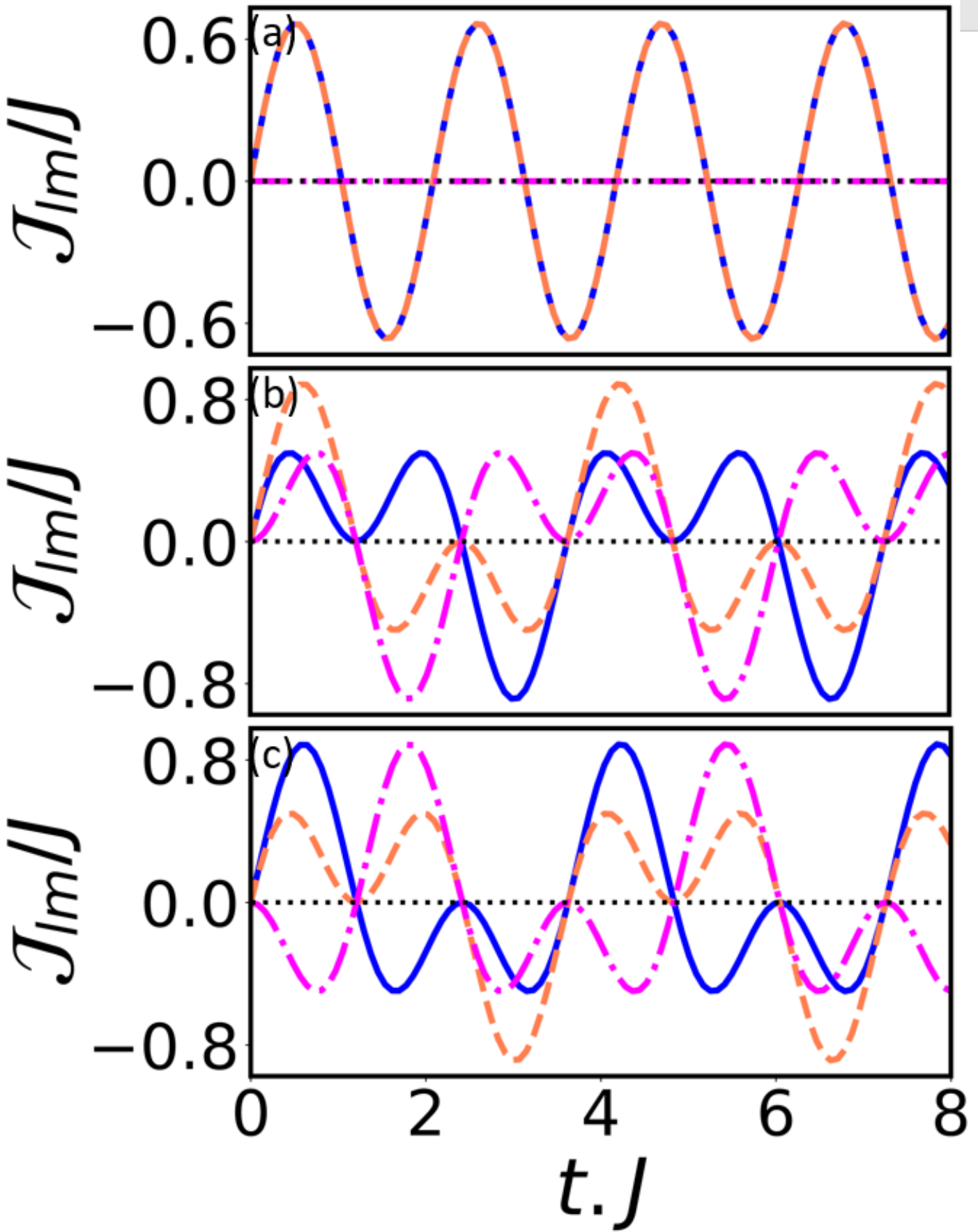}
\caption{Local currents for closed system ($J/\omega=3\times10^{-3}$, $\gamma=0$): (a),(b) and (c) show the local currents $\mathcal{J}_{21}$ (blue-solid), $\mathcal{J}_{23}$ (orange-dashed) and $\mathcal{J}_{13}$ (magenta-dash dotted) as a function of time $t$ for the initial state $|010\rangle$. }  
\label{fig:2}    
\end{center}
\end{figure}

However, a description of such type restricts us from exploring the strong coupling regime where $J>\gamma$. To this end, we use an exact numeric approach which has been recently proposed in Ref~\cite{rivas2010markovian}. Such an approach essentially models the quantum bath as a collection of a large, yet finite number of harmonic oscillators. In what follows, we write the composite system-bath model as: 
\begin{align}\label{Ham_total}
        &\emph{$\tilde{H}$} = \emph{H} +  \sum_{\alpha=c,h} \left[ \emph{H}_{\alpha}+\emph{V}_{\alpha}\right] \nonumber\\
        &\emph{H}_{\alpha} = \sum_{j}\omega_{\alpha, j}b_{\alpha, j}^\dagger b_{\alpha, j},\emph{V}_{h} = \sum_{j}g_{h, j}\left(b_{h, j}^\dagger a_2 + a_2^\dagger b_{h, j}\right),\nonumber\\
        &\emph{V}_{c} = \sum_{j}g_{c, j}\left(b_{c, j}^\dagger (a_1 + a_3) + (a_1^\dagger +  a_3^\dagger) b_{c, j}\right),  
\end{align}
where $b_{\alpha, j}$ $b_{\alpha, j}^\dagger$ denote annihilation (creation) operators of the baths, $\omega_{\alpha, j}$ are the frequencies corresponding to the bath modes, and $g_{\alpha, j}$ denote the interaction strength between system and bath modes.
where we have considered the RWA part in the system bath coupling Hamiltonian as $\omega\gg {\rm max}({g_{\alpha,j}}) $. The dynamics of the combined system-bath model can then be exactly given by the Heisenberg equation of motion. In a compact form, with the notion of 
$\mathcal{A} = (a_2, b_{h, 1}, b_{h,2}, \ldots, b_{h,N}, a_1, a_3, b_{c, 1}, b_{c,2}, \ldots, b_{c,N})^T$ and $\mathcal{A}^{\dagger} = (a_2^{\dagger}, b_{h, 1}^{\dagger}, b_{h,2}^{\dagger}, \ldots, b_{h,N}^{\dagger}, a_1^{\dagger}, a_3^{\dagger}, b_{c, 1}^{\dagger}, b_{c,2}^{\dagger}, \ldots, b_{c,N}^{\dagger})^T$, the time evolution of combined state-variables is given by,
\begin{align}
&\mathcal{A}(t) = e^{-iWt}\mathcal{A}(0),\\ 
&\mathcal{A}^\dagger(t) = e^{i\overline{W}t}\mathcal{A}^\dagger(0),
\end{align}
where $\overline{W}$ is the conjugate of the complex matrix $W$. (See the supplemental material for subsequent discussion.) Note that a description of such type essentially implies a unitary evolution of the composite system-bath model. The evaluation of the local currents then involves an estimation of the intrasystem correlation involving the $1$st, $(N+2)$'th, and the $(N+3$)'th element of the covariance matrix.  

\begin{figure}[!t]
\begin{center}
\includegraphics[width=6 cm]{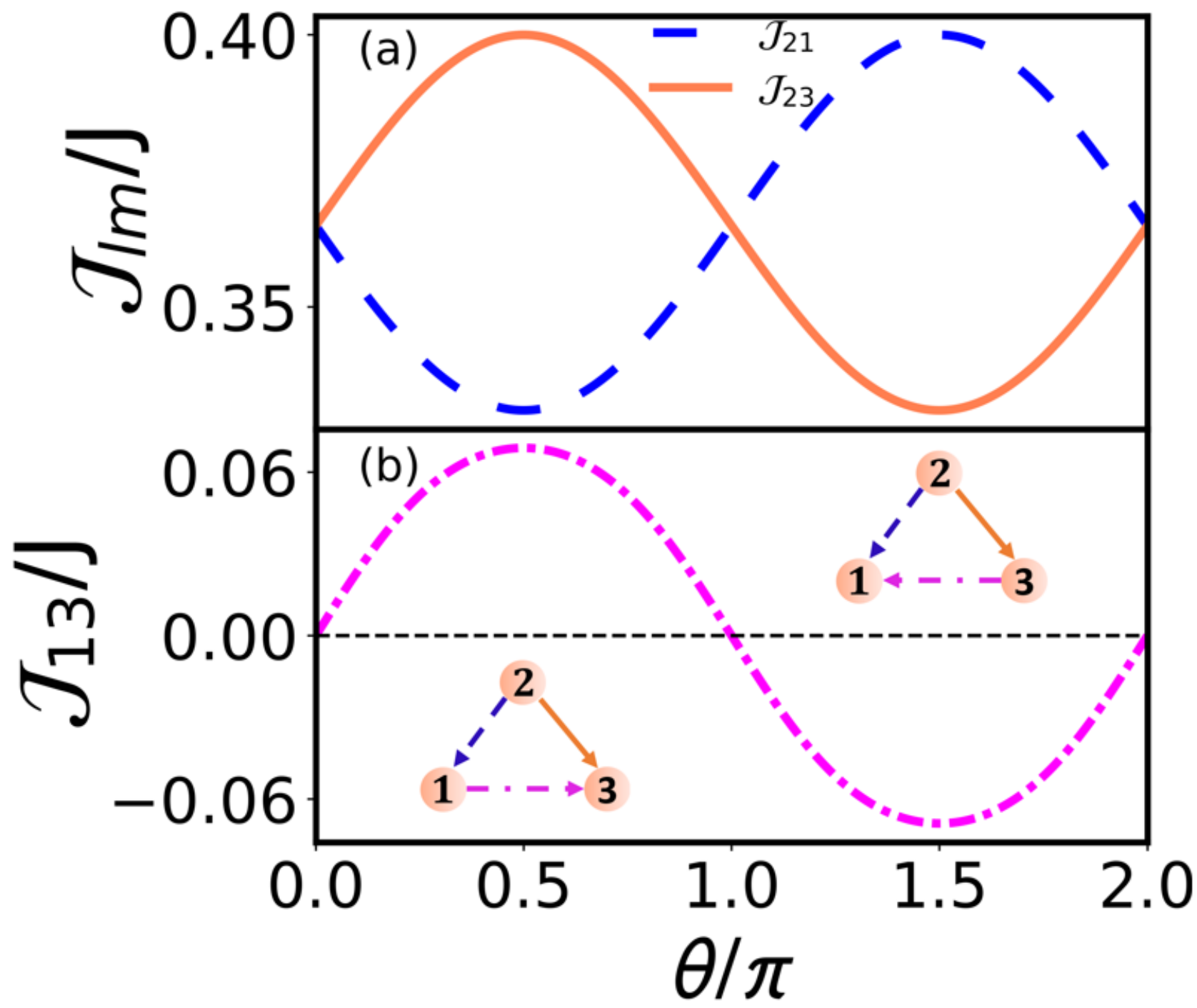}
\caption{Steadystate local heat currents ($J/\gamma=0.1$, $\gamma/\omega=0.03$): (a)$\mathcal{J}_{21}$ (blue-solid),  $\mathcal{J}_{23}$ (orange-dashed) and (b)$\mathcal{J}_{13}$ (magenta-dash dotted) in units of $J$ as a function of $\theta$. $\mathcal{J}_{13}$ at $\theta=\pi/2$ and $3\pi/2$ show distinct patterns of flow.}
\label{fig:3}    
\end{center}
\end{figure}

{\it Results:} First, we consider the unitary behaviour where no thermal baths are attached to the system. For the simulation of the closed system, we use $J/\omega=3\times10^{-3}$ and initiate the system at $|010\rangle$ to observe its dynamics for different $\theta$. The local currents are depicted in Fig.~\ref{fig:2}. It is seen that for $\theta = n \pi$, ($n$ being an integer) $\mathcal{J}_{21}$ and $\mathcal{J}_{23}$ are the only two currents that flow across the system, with equal amplitudes. For $\theta \neq n \pi$, a nonzero current $\mathcal{J}_{13}$ emerges, with its direction being depended on $\theta$. For instances, at $Jt=2$ we see a flow $1\leftarrow3$ at $\theta = \pi/2$, while at $\theta = 3 \pi/2$ the flow is $1\rightarrow3$. It is worthwhile to note that, for $\theta\in[0,2\pi]$ we do not observe a full circulation of the local currents and no particular pattern of flow of internal current is persistent under unitary dynamics.

We now turn our discussion to study the nonequilibrium aspects of our system. To this end, we study the weak coupling regime first, where the Lindblad Quantum Master Equation (LQME) holds good. Fig.~\ref{fig:3} depicts the local thermal currents as a function of the phase $\theta$. It is shown that when in steady-state, due to the chosen temperature asymmetry $T_H/\omega = 5, T_C/\omega = 3$, two heat currents flow simultaneously from the site $2\rightarrow1$ and $2\rightarrow3$. We find these currents to be perfectly symmetric at $\theta=n\pi$, while at any other $\theta\neq n\pi$, the symmetry gets broken, giving rise to an atypical current flowing between the sites $1$ and $3$. It may seem that this atypical current vioaltes the zeroth law of thermodyanmics in which any flow if forbidden between two thermally equilibrated bodies. Howver, we will discuss the thermodynamic consistency of such flow in the subsequent section.  What is more here is the path taken by the current, \textit{viz.} $1\rightarrow3$ and $3\rightarrow1$ respectively for $0<\theta<\pi$ and $\pi<\theta<2\pi$. Such an atypical behaviour of $\langle \mathcal{J}_{13} \rangle$ (for $J/\gamma<1 $) can be well captured by an approximate closed-form solution $\langle\mathcal{J}_{13}\rangle \approx 
 2J \left(N_H - N_C\right) \left(J/\gamma\right)^2  \sin{\theta}$. One can then infer that increasing the temperature gradient increases the current amplitude and reversing the bath configuration changes the current direction. The overall pattern of the internal current circulation can be categorized into three distinct regimes of flow, \textit{viz} (\textit{i}) anti-parallel currents $1\leftarrow2\rightarrow3$ at $\theta=n\pi$ ($n=0,1,2$), and two parallel currents (\textit{ii})  $2\rightarrow3,2\rightarrow1\rightarrow3$ at $\theta=\pi/2$, and (\textit{iii}) $2\rightarrow1,2\rightarrow3\rightarrow1$ at $\theta=3\pi/2$.

To get into the physics of such atypical current, we consider an interferometric approach where we introduce two \enquote{hybrid} modes $A_{+}=\left(a_1 + e^{i\theta}a_3\right)/\sqrt{2}$ and $A_{-}=\left(a_3 - e^{-i\theta}a_1\right)/\sqrt{2}$ and rewrite the Hamiltonian \eqref{Ham} as 
\begin{align}
    \tilde{H}_{hyb}=&\omega a_2 ^\dagger a_2 + \sum_{l=\pm} \omega_l A_l^\dagger A_l + \left( a_2^\dagger \sum_{l=\pm} J_l A_l + h.c \right).
\end{align}
where $\omega_\pm=\omega\pm J$ are the resonance frequencies and $J_\pm=J\left(1\pm e^{\mp i\theta}\right)/\sqrt{2}$ are the coupling strengths. The corresponding Lindblad superoperators when expressed in the hybrid mode bases $ L_l = \{a_2, A_+, A_-\}$ provides insight into the interaction of the hybrid system with the baths. In a similar spirit,  we recast the current expression as $ \langle \mathcal{J}_{13} \rangle = i J \left(e^{i \theta}\langle A^\dagger_{+}A_- \rangle - e^{-i \theta} \langle A^\dagger_- A_+\rangle\right)$. When $\theta=n\pi$, site $2$ gets decoupled from one of the hybrid modes $A_+$ (for odd $n$) and $A_-$ (for even $n$), which we call as the \textit{dark mode}. For any other $\theta\neq n\pi$ the \textit{dark mode} disappears, resulting in a coherent interaction between $A_+$ and $A_-$. In other words, via dark mode breaking, the two hybrid modes constructively interfere, resulting in a finite nonequilibrium current, with its direction being dependent on $\theta$.  We also note that the symmetric choice of system bath parameters $(J,\gamma)$ does not give rise to any incoherent coupling between $A_+$ and $A_-$, through the common reservoir of sites $1$ and $3$. Nevertheless, any asymmetric choice of such parameters would lead to a non-zero current between $1\leftrightarrow3$ which has been discussed in Supplemental material.

\begin{figure}[!t]
\begin{center}
   \includegraphics[width = 8 cm]{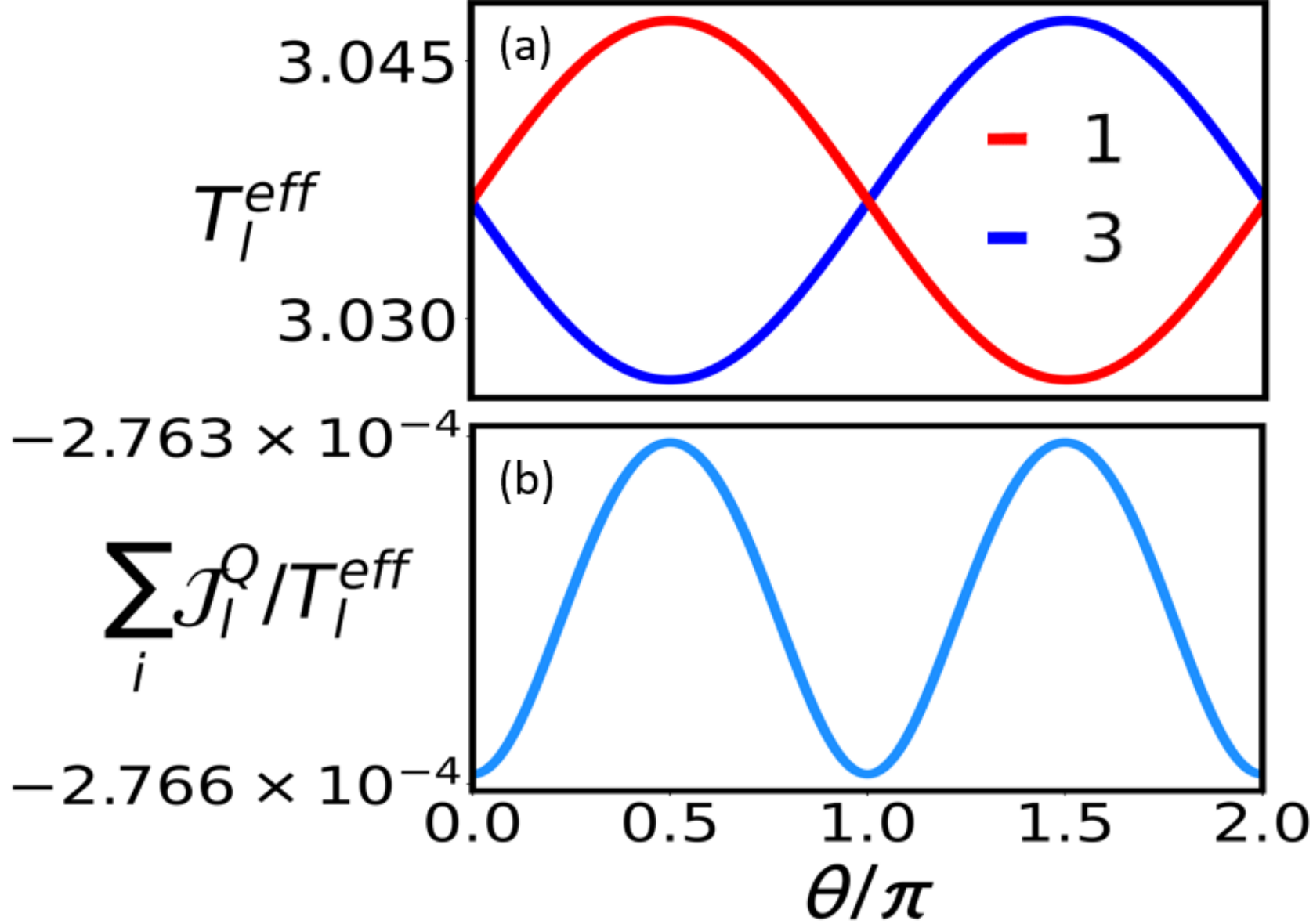}
   \caption{ (a) Effective steady-state temperature of sites 1 and 3 and (b) $\Sigma_{l}\mathcal{J}^{Q}_{l}/T^{eff}_{l}$ as a function of phase $\theta$ for parameters $J/\gamma=0.1$, $\gamma/\omega=0.03$, $T_{H}/\omega=5$ and $T_{C}/\omega=3$}
  \label{fig:4}  
 \end{center}
  \end{figure}
We next turn to the thermodynamic consistency of our findings from LQME and its implications. We assume that, in long time, each site $l$ evolves to a thermal steady-state, characterized by an effective temperate $T^{eff}_{l}$. An estimation of these temperatures can be made by minimizing the trace distance $D(\rho^{NESS}_l,\rho^{th}_l)\equiv \frac{1}{2}{\parallel \rho^{NESS}_{l} - \rho^{th}_l \parallel}$ between the reduced state $\rho^{NESS}_{l}=Tr_{\neq l} \rho^{NESS}$ and an effective thermal state $\rho^{th}_l =\mathrm{exp}\left(-\frac{\omega a_l ^\dagger a_l }{T^{eff}_l} \right)/Tr\left[\mathrm{exp}\left(-\frac{\omega a_l ^\dagger a_l }{T^{eff}_l}\right)\right]$ density matrices. Fig.~\ref{fig:4}(a) depicts the temperatures $T^{eff}_l$ against the phase value $\theta$. We find that when the dark mode sets in, the sites $1$ and $3$ attain local thermal equilibrium, as characterized by the same effective temperature ($T^{eff}_1=T^{eff}_3$). While breaking this dark mode results in a temperature gradient $T^{eff}_1<T^{eff}_3$ and  $T^{eff}_3<T^{eff}_1$ respectively for $0<\theta<\pi$ and $\pi<\theta<2\pi$. This also provides an intuitive explanation for the reversal of current between sites $1$ and $3$. Next, to ascertain the validity of the second law of thermodynamics, we check for the inequality~\cite{kosloff2013quantum}
\begin{equation}
  \sum_{l} \frac{\mathcal{J}^{Q}_{l}}{T^{eff}_{l}}\leq 0.   
\end{equation}
To be consistent with the local description of the quantum master equation~\cite{landi2022nonequilibrium}, we define the energy current operator (bath$\rightarrow$ site) $\mathcal{J}^{Q}_{l}= \text{Tr}\{H_{l}\mathcal{D}_{l}(\rho^{\text{NESS}})\}$, in terms of the local Hamiltonian $H_{l}=\omega a^{\dagger}_l a_l$. From Fig.~\ref{fig:4}(b), we find the inequality holds for all $\theta$, validating the second law of thermodynamics. To capture the variation with respect to $\theta$, we take into account the \textit{entropy production rate}, defined as $\dot{\Sigma} = -\sum_{l}\mathcal{J}^{Q}_{l}/T^{eff}_{l} $. Fig.~\ref{fig:4}(b) shows that $\dot{\Sigma}$ becomes maximum and minimum respectively at $\theta = n\pi$ and $\theta = (2n+1)\frac{\pi}{2}$. Remarkably, in the case where the environment consists of thermal baths, the entropy production is exactly (up to a difference in sign) the correlations shared between the system and environment~\cite{deffner2019quantum}.

  \begin{figure}[!t]
\begin{center}
   \includegraphics[width = 8 cm]{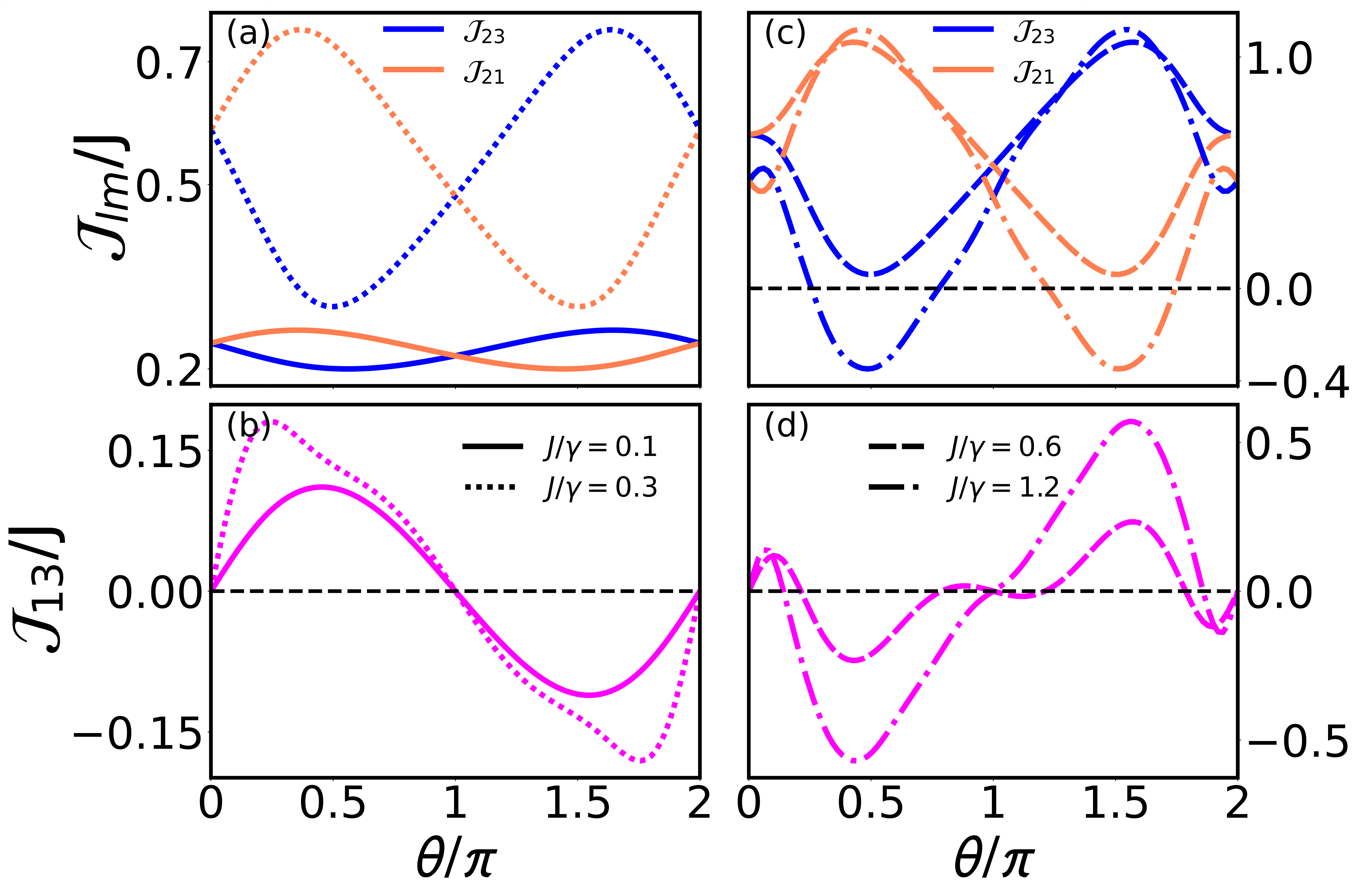}
   \caption{ (a) Steadystate local currents through an exact numerics simulation, for different $J/\gamma$ ratios. The solid, dotted, dashed and dashed-dotted lines respectively correspond to $J/\gamma=0.1$, 0.3, 0.6 and 1.2. Common parameters, $\gamma/\omega=0.03$, $T_{H}/\omega=5$ and $T_{C}/\omega=3$}
  \label{fig:5}  
 \end{center}
  \end{figure}
  
Finally, we focus on the strong coupling regime. Fig.~\ref{fig:5} depicts the local steadystate currents across the sites, obtained through exact numerics simulations. To benchmark our findings, we first consider small $J/\gamma$ ratios. As expected, we find a good degree of qualitative agreement between the LQME and the exact simulation. Increasing $J/\gamma$ further, results in a departure from the conventional behaviour. In particular, there emereges a critical $(J/\gamma)_{\rm crit}$ ratio where the the atypical current $\mathcal{J}_{13}(\pi/2)$ vanishes (Fig.~\ref{fig:5}). While above $(J/\gamma)_{\rm crit}$, the current undergoes a $\pi$ phase-shift, changing from $\mathcal{J}_{13}\sim \sin{\theta}$ to $\mathcal{J}_{13}\sim \sin(\theta+\pi)$. Notably, such $\pi$ phase transition in the current-phase relation has been observed in \enquote{electronic} three-terminal devices, including molecular Andreev interferometers \cite{plaszko2020quantum}, superconductor-normal-metal-superconductor Josephson Junctions \cite{baselmans1999reversing, baselmans2002direct}. Remarkably, to the best of our knowledge, such $\pi$ phase transition in nonequilibrium current has not been observed before. It is also interesting to note here that at $J/\gamma = 1.2$ the currents $\mathcal{J}_{23}(\pi/2)$ ($\mathcal{J}_{21}(3\pi/2)$) becomes negative implying a flow $3\rightarrow2$ ($1\rightarrow2$), or in other words from cold to the hot bath. However, if the total current $\mathcal{J}_T = \mathcal{J}_{21} + \mathcal{J}_{23}$ is considered, we find that $\mathcal{J}_T$ remains positive throughout $\theta \in [0,2\pi]$, maintaining the thermodynamic consistency of the flow.

\begin{figure}[!t]
\begin{center}
   \includegraphics[width = 7.5 cm]{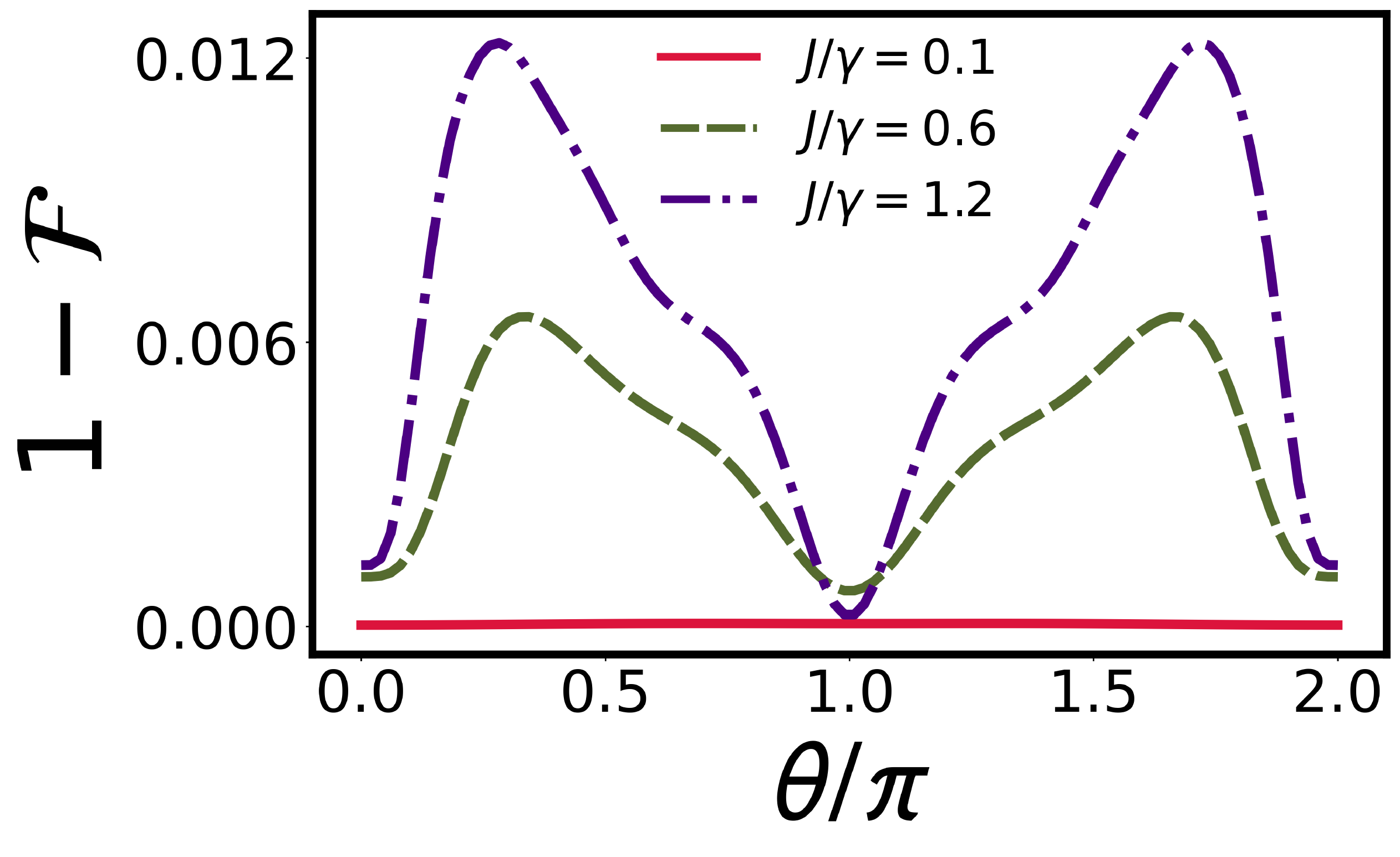}
   \caption{Fidelity between the states of subsystem 1 and 3 from exact numerics and LQME for different $J/\gamma$ ratios. The solid, dashed and dash dotted lines respectively correspond to $J/\gamma = 0.1, 0.6$ and $1.2$. with $\gamma/\omega=0.03$, $T_{H}/\omega=5$ and $T_{C}/\omega=3$}
  \label{fig:6}  
 \end{center}
\end{figure}
Notably, the outcomes derived from LQME do not manifest any critical behaviour. This discrepancy can be ascribed to the rapid divergence between the intrasystem correlations obtained from Exact numerics and those of the LQME, as previously observed \cite{hofer2017markovian}. Furthermore, prior investigations have scrutinized the fidelity of states evolved by LQME and exact numerics across varying $J/\gamma$ ratios, revealing an anticipated decrease in fidelity for higher ratios. Therefore, exploring such phenomena in systems augmented by complex phases for smaller $J/\gamma$ ratios presents an intriguing avenue. To this end, we analyze the fidelity of subsystem (1 and 3) states described by two-mode Gaussian states \cite{marian2012uhlmann}, derived from LQME and exact numerics as a function of $\theta$ for different $J/\gamma$ ratios. As shown in Fig.~\ref{fig:6}, for $J/\gamma=0.1$, the states exhibit high fidelity across all $\theta$, whereas for $J/\gamma = 0.6$ and $1.2$, fidelity diminishes for $\theta \neq n\pi$. This observation suggests a divergence in the descriptions from LQME and exact numerics for complex phase-dependent couplings, even at lower $J/\gamma$ ratios.

\begin{figure}[!t]
\begin{center}
   \includegraphics[width = 8 cm]{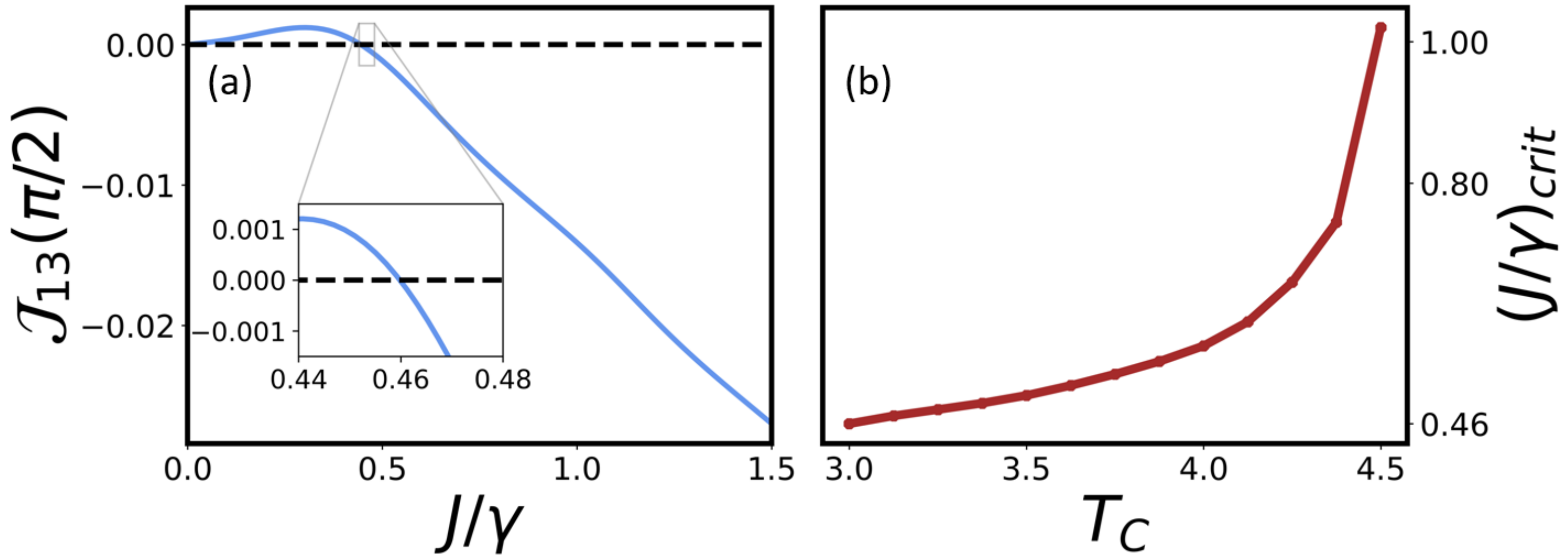}
   \caption{ (a) Internal current $\mathcal{J}_{13}$ at $\theta=\pi/2$ and $T_{C}/\omega=3$ as a function of $J/\gamma$. (b) $(J/\gamma)_{\rm crit}$ with respect to the cold bath temperatures $T_C$. Common parameters, $\gamma/\omega=0.03$, $T_{H}/\omega=5$ }
  \label{fig:7}  
 \end{center}
\end{figure}

To probe the critical ratio $(J/\gamma)_{\rm crit}$ where the phase reversal occurs, we consider $\mathcal{J}_{13}(\pi/2)$ as a figure of merit and plot it for $J/\gamma$. Fig.~\ref{fig:7}(a) depicts that for the chosen temperature bias, the transition occurs at $(J/\gamma)_{\rm crit} = 0.46$, and, beyond $(J/\gamma)_{\rm crit}$, $\mathcal{J}_{13}(\pi/2)$ becomes negative ($3\leftarrow2$), with monotonically increasing amplitude. Here, it is also interesting to look at how the critical ratio depends on the bath temperature gradient. From Fig.~\ref{fig:7}(b) it can be seen that as we lower the temperature gradient (increase $T_C$), $(J/\gamma)_{\rm crit}$ shifts towards a higher $J/\gamma$ value. In particular, for $T_C = 4.5$ the phase reversal occurs at $(J/\gamma)_{\rm crit}\approx 1$. We can thus safely remark that such $\pi$ phase transition in nonequilibrium nonreciprocal current depends on both the $(J/\gamma)$ value and the temperature bias, under which the system operates.

{\it Conclusion:} We have studied thermal transport in a quantum-engineered bosonic trimer. When in contact with two heat reservoirs, the system exhibits a nonequilibrium nonreciprocal current between two bodies, originally in thermal equilibrium. The physics of such an atypical current is explained through the phenomenon of dark-mode breaking, with symmetrically chosen system-bath parameters. Further investigation into the strong coupling regime, reveals a novel phase reversal where the current changes from $\mathcal{J}\sim \sin{\theta}$ to $\mathcal{J}\sim \sin(\theta+\pi)$. 

{\it Outlook:} Our findings may find its application in the development of novel phonon electronic devices such as heat switches or phononic gates. For instance, a \textit{swap} device, with its action defined as $\mathcal{J}_{lm}=\mathcal{J}_{ml}$. Fig.~\ref{fig:3}(a) shows that such an action can be realized by tuning $\theta$ as $\mathcal{J}_{21}(\theta) = \mathcal{J}_{23}(\theta + \pi)$. Moreover, one could also consider a three-terminal thermal \textit{switch}. Analogous to an electrical switch, a thermal switch will then operate to block (switch \textit{off}) or allow (switch \textit{on}) the heat current to flow between two terminals. Fig.~\ref{fig:3}(b) shows how one can turn on and off the device and manipulate the flow by tuning $\theta$, respectively, in the interval of $[0,\pi]$ and $[\pi,2\pi]$. It is important to note that such actions fail for intermediate $J/\gamma$ values where the nonequilibrium nonreciprocal current undergoes a phase reversal. The results presented here are not particular to any physical system and thus can be explored further for various dark mode assisted transport properties.  

{\it Acknowledgement:}
GC and SC gratefully acknowledge H.S. Dhar for his valuable suggestions. MM acknowledges the support of this research by the National Research Foundation, Singapore and A*STAR under its Quantum Engineering Programme (NRF2021-QEP2-02-P10 and NRF2021-QEP2-03-P07).



\begin{thebibliography}{33}%
\makeatletter
\providecommand \@ifxundefined [1]{%
 \@ifx{#1\undefined}
}%
\providecommand \@ifnum [1]{%
 \ifnum #1\expandafter \@firstoftwo
 \else \expandafter \@secondoftwo
 \fi
}%
\providecommand \@ifx [1]{%
 \ifx #1\expandafter \@firstoftwo
 \else \expandafter \@secondoftwo
 \fi
}%
\providecommand \natexlab [1]{#1}%
\providecommand \enquote  [1]{``#1''}%
\providecommand \bibnamefont  [1]{#1}%
\providecommand \bibfnamefont [1]{#1}%
\providecommand \citenamefont [1]{#1}%
\providecommand \href@noop [0]{\@secondoftwo}%
\providecommand \href [0]{\begingroup \@sanitize@url \@href}%
\providecommand \@href[1]{\@@startlink{#1}\@@href}%
\providecommand \@@href[1]{\endgroup#1\@@endlink}%
\providecommand \@sanitize@url [0]{\catcode `\\12\catcode `\$12\catcode `\&12\catcode `\#12\catcode `\^12\catcode `\_12\catcode `\%12\relax}%
\providecommand \@@startlink[1]{}%
\providecommand \@@endlink[0]{}%
\providecommand \url  [0]{\begingroup\@sanitize@url \@url }%
\providecommand \@url [1]{\endgroup\@href {#1}{\urlprefix }}%
\providecommand \urlprefix  [0]{URL }%
\providecommand \Eprint [0]{\href }%
\providecommand \doibase [0]{https://doi.org/}%
\providecommand \selectlanguage [0]{\@gobble}%
\providecommand \bibinfo  [0]{\@secondoftwo}%
\providecommand \bibfield  [0]{\@secondoftwo}%
\providecommand \translation [1]{[#1]}%
\providecommand \BibitemOpen [0]{}%
\providecommand \bibitemStop [0]{}%
\providecommand \bibitemNoStop [0]{.\EOS\space}%
\providecommand \EOS [0]{\spacefactor3000\relax}%
\providecommand \BibitemShut  [1]{\csname bibitem#1\endcsname}%
\let\auto@bib@innerbib\@empty
\bibitem [{\citenamefont {Vinjanampathy}\ and\ \citenamefont {Anders}(2016)}]{vinjanampathy2016quantum}%
  \BibitemOpen
  \bibfield  {author} {\bibinfo {author} {\bibfnamefont {S.}~\bibnamefont {Vinjanampathy}}\ and\ \bibinfo {author} {\bibfnamefont {J.}~\bibnamefont {Anders}},\ }\href@noop {} {\bibfield  {journal} {\bibinfo  {journal} {Contemporary Physics}\ }\textbf {\bibinfo {volume} {57}},\ \bibinfo {pages} {545} (\bibinfo {year} {2016})}\BibitemShut {NoStop}%
\bibitem [{\citenamefont {Kosloff}(2013)}]{kosloff2013quantum}%
  \BibitemOpen
  \bibfield  {author} {\bibinfo {author} {\bibfnamefont {R.}~\bibnamefont {Kosloff}},\ }\href@noop {} {\bibfield  {journal} {\bibinfo  {journal} {Entropy}\ }\textbf {\bibinfo {volume} {15}},\ \bibinfo {pages} {2100} (\bibinfo {year} {2013})}\BibitemShut {NoStop}%
\bibitem [{\citenamefont {Bouton}\ \emph {et~al.}(2021)\citenamefont {Bouton}, \citenamefont {Nettersheim}, \citenamefont {Burgardt}, \citenamefont {Adam}, \citenamefont {Lutz},\ and\ \citenamefont {Widera}}]{bouton2021quantum}%
  \BibitemOpen
  \bibfield  {author} {\bibinfo {author} {\bibfnamefont {Q.}~\bibnamefont {Bouton}}, \bibinfo {author} {\bibfnamefont {J.}~\bibnamefont {Nettersheim}}, \bibinfo {author} {\bibfnamefont {S.}~\bibnamefont {Burgardt}}, \bibinfo {author} {\bibfnamefont {D.}~\bibnamefont {Adam}}, \bibinfo {author} {\bibfnamefont {E.}~\bibnamefont {Lutz}},\ and\ \bibinfo {author} {\bibfnamefont {A.}~\bibnamefont {Widera}},\ }\href@noop {} {\bibfield  {journal} {\bibinfo  {journal} {Nature Communications}\ }\textbf {\bibinfo {volume} {12}},\ \bibinfo {pages} {2063} (\bibinfo {year} {2021})}\BibitemShut {NoStop}%
\bibitem [{\citenamefont {Levy}\ \emph {et~al.}(2012)\citenamefont {Levy}, \citenamefont {Alicki},\ and\ \citenamefont {Kosloff}}]{levy2012quantum}%
  \BibitemOpen
  \bibfield  {author} {\bibinfo {author} {\bibfnamefont {A.}~\bibnamefont {Levy}}, \bibinfo {author} {\bibfnamefont {R.}~\bibnamefont {Alicki}},\ and\ \bibinfo {author} {\bibfnamefont {R.}~\bibnamefont {Kosloff}},\ }\href@noop {} {\bibfield  {journal} {\bibinfo  {journal} {Physical Review E}\ }\textbf {\bibinfo {volume} {85}},\ \bibinfo {pages} {061126} (\bibinfo {year} {2012})}\BibitemShut {NoStop}%
\bibitem [{\citenamefont {Ronzani}\ \emph {et~al.}(2018)\citenamefont {Ronzani}, \citenamefont {Karimi}, \citenamefont {Senior}, \citenamefont {Chang}, \citenamefont {Peltonen}, \citenamefont {Chen},\ and\ \citenamefont {Pekola}}]{ronzani2018tunable}%
  \BibitemOpen
  \bibfield  {author} {\bibinfo {author} {\bibfnamefont {A.}~\bibnamefont {Ronzani}}, \bibinfo {author} {\bibfnamefont {B.}~\bibnamefont {Karimi}}, \bibinfo {author} {\bibfnamefont {J.}~\bibnamefont {Senior}}, \bibinfo {author} {\bibfnamefont {Y.-C.}\ \bibnamefont {Chang}}, \bibinfo {author} {\bibfnamefont {J.~T.}\ \bibnamefont {Peltonen}}, \bibinfo {author} {\bibfnamefont {C.}~\bibnamefont {Chen}},\ and\ \bibinfo {author} {\bibfnamefont {J.~P.}\ \bibnamefont {Pekola}},\ }\href@noop {} {\bibfield  {journal} {\bibinfo  {journal} {Nature Physics}\ }\textbf {\bibinfo {volume} {14}},\ \bibinfo {pages} {991} (\bibinfo {year} {2018})}\BibitemShut {NoStop}%
\bibitem [{\citenamefont {Senior}\ \emph {et~al.}(2020)\citenamefont {Senior}, \citenamefont {Gubaydullin}, \citenamefont {Karimi}, \citenamefont {Peltonen}, \citenamefont {Ankerhold},\ and\ \citenamefont {Pekola}}]{senior2020heat}%
  \BibitemOpen
  \bibfield  {author} {\bibinfo {author} {\bibfnamefont {J.}~\bibnamefont {Senior}}, \bibinfo {author} {\bibfnamefont {A.}~\bibnamefont {Gubaydullin}}, \bibinfo {author} {\bibfnamefont {B.}~\bibnamefont {Karimi}}, \bibinfo {author} {\bibfnamefont {J.~T.}\ \bibnamefont {Peltonen}}, \bibinfo {author} {\bibfnamefont {J.}~\bibnamefont {Ankerhold}},\ and\ \bibinfo {author} {\bibfnamefont {J.~P.}\ \bibnamefont {Pekola}},\ }\href@noop {} {\bibfield  {journal} {\bibinfo  {journal} {Communications Physics}\ }\textbf {\bibinfo {volume} {3}},\ \bibinfo {pages} {40} (\bibinfo {year} {2020})}\BibitemShut {NoStop}%
\bibitem [{\citenamefont {Mandarino}\ \emph {et~al.}(2021)\citenamefont {Mandarino}, \citenamefont {Joulain}, \citenamefont {G{\'o}mez},\ and\ \citenamefont {Bellomo}}]{mandarino2021thermal}%
  \BibitemOpen
  \bibfield  {author} {\bibinfo {author} {\bibfnamefont {A.}~\bibnamefont {Mandarino}}, \bibinfo {author} {\bibfnamefont {K.}~\bibnamefont {Joulain}}, \bibinfo {author} {\bibfnamefont {M.~D.}\ \bibnamefont {G{\'o}mez}},\ and\ \bibinfo {author} {\bibfnamefont {B.}~\bibnamefont {Bellomo}},\ }\href@noop {} {\bibfield  {journal} {\bibinfo  {journal} {Physical Review Applied}\ }\textbf {\bibinfo {volume} {16}},\ \bibinfo {pages} {034026} (\bibinfo {year} {2021})}\BibitemShut {NoStop}%
\bibitem [{\citenamefont {S{\"a}{\"a}skilahti}\ \emph {et~al.}(2013)\citenamefont {S{\"a}{\"a}skilahti}, \citenamefont {Oksanen},\ and\ \citenamefont {Tulkki}}]{saaskilahti2013thermal}%
  \BibitemOpen
  \bibfield  {author} {\bibinfo {author} {\bibfnamefont {K.}~\bibnamefont {S{\"a}{\"a}skilahti}}, \bibinfo {author} {\bibfnamefont {J.}~\bibnamefont {Oksanen}},\ and\ \bibinfo {author} {\bibfnamefont {J.}~\bibnamefont {Tulkki}},\ }\href@noop {} {\bibfield  {journal} {\bibinfo  {journal} {Physical Review E}\ }\textbf {\bibinfo {volume} {88}},\ \bibinfo {pages} {012128} (\bibinfo {year} {2013})}\BibitemShut {NoStop}%
\bibitem [{\citenamefont {{\v{S}}afr{\'a}nek}\ \emph {et~al.}(2023)\citenamefont {{\v{S}}afr{\'a}nek}, \citenamefont {Rosa},\ and\ \citenamefont {Binder}}]{vsafranek2023work}%
  \BibitemOpen
  \bibfield  {author} {\bibinfo {author} {\bibfnamefont {D.}~\bibnamefont {{\v{S}}afr{\'a}nek}}, \bibinfo {author} {\bibfnamefont {D.}~\bibnamefont {Rosa}},\ and\ \bibinfo {author} {\bibfnamefont {F.~C.}\ \bibnamefont {Binder}},\ }\href@noop {} {\bibfield  {journal} {\bibinfo  {journal} {Physical Review Letters}\ }\textbf {\bibinfo {volume} {130}},\ \bibinfo {pages} {210401} (\bibinfo {year} {2023})}\BibitemShut {NoStop}%
\bibitem [{\citenamefont {Levy}\ and\ \citenamefont {Kosloff}(2014)}]{levy2014local}%
  \BibitemOpen
  \bibfield  {author} {\bibinfo {author} {\bibfnamefont {A.}~\bibnamefont {Levy}}\ and\ \bibinfo {author} {\bibfnamefont {R.}~\bibnamefont {Kosloff}},\ }\href@noop {} {\bibfield  {journal} {\bibinfo  {journal} {Europhysics Letters}\ }\textbf {\bibinfo {volume} {107}},\ \bibinfo {pages} {20004} (\bibinfo {year} {2014})}\BibitemShut {NoStop}%
\bibitem [{\citenamefont {Rivas}\ \emph {et~al.}(2010)\citenamefont {Rivas}, \citenamefont {Plato}, \citenamefont {Huelga},\ and\ \citenamefont {Plenio}}]{rivas2010markovian}%
  \BibitemOpen
  \bibfield  {author} {\bibinfo {author} {\bibfnamefont {A.}~\bibnamefont {Rivas}}, \bibinfo {author} {\bibfnamefont {A.~D.~K.}\ \bibnamefont {Plato}}, \bibinfo {author} {\bibfnamefont {S.~F.}\ \bibnamefont {Huelga}},\ and\ \bibinfo {author} {\bibfnamefont {M.~B.}\ \bibnamefont {Plenio}},\ }\href@noop {} {\bibfield  {journal} {\bibinfo  {journal} {New Journal of Physics}\ }\textbf {\bibinfo {volume} {12}},\ \bibinfo {pages} {113032} (\bibinfo {year} {2010})}\BibitemShut {NoStop}%
\bibitem [{\citenamefont {Cattaneo}\ \emph {et~al.}(2019)\citenamefont {Cattaneo}, \citenamefont {Giorgi}, \citenamefont {Maniscalco},\ and\ \citenamefont {Zambrini}}]{cattaneo2019local}%
  \BibitemOpen
  \bibfield  {author} {\bibinfo {author} {\bibfnamefont {M.}~\bibnamefont {Cattaneo}}, \bibinfo {author} {\bibfnamefont {G.~L.}\ \bibnamefont {Giorgi}}, \bibinfo {author} {\bibfnamefont {S.}~\bibnamefont {Maniscalco}},\ and\ \bibinfo {author} {\bibfnamefont {R.}~\bibnamefont {Zambrini}},\ }\href@noop {} {\bibfield  {journal} {\bibinfo  {journal} {New Journal of Physics}\ }\textbf {\bibinfo {volume} {21}},\ \bibinfo {pages} {113045} (\bibinfo {year} {2019})}\BibitemShut {NoStop}%
\bibitem [{\citenamefont {De~Chiara}\ \emph {et~al.}(2018)\citenamefont {De~Chiara}, \citenamefont {Landi}, \citenamefont {Hewgill}, \citenamefont {Reid}, \citenamefont {Ferraro}, \citenamefont {Roncaglia},\ and\ \citenamefont {Antezza}}]{de2018reconciliation}%
  \BibitemOpen
  \bibfield  {author} {\bibinfo {author} {\bibfnamefont {G.}~\bibnamefont {De~Chiara}}, \bibinfo {author} {\bibfnamefont {G.}~\bibnamefont {Landi}}, \bibinfo {author} {\bibfnamefont {A.}~\bibnamefont {Hewgill}}, \bibinfo {author} {\bibfnamefont {B.}~\bibnamefont {Reid}}, \bibinfo {author} {\bibfnamefont {A.}~\bibnamefont {Ferraro}}, \bibinfo {author} {\bibfnamefont {A.~J.}\ \bibnamefont {Roncaglia}},\ and\ \bibinfo {author} {\bibfnamefont {M.}~\bibnamefont {Antezza}},\ }\href@noop {} {\bibfield  {journal} {\bibinfo  {journal} {New Journal of Physics}\ }\textbf {\bibinfo {volume} {20}},\ \bibinfo {pages} {113024} (\bibinfo {year} {2018})}\BibitemShut {NoStop}%
\bibitem [{\citenamefont {Lai}\ \emph {et~al.}(2018)\citenamefont {Lai}, \citenamefont {Di~Ventra}, \citenamefont {Scheibner},\ and\ \citenamefont {Chien}}]{lai2018tunable}%
  \BibitemOpen
  \bibfield  {author} {\bibinfo {author} {\bibfnamefont {C.-Y.}\ \bibnamefont {Lai}}, \bibinfo {author} {\bibfnamefont {M.}~\bibnamefont {Di~Ventra}}, \bibinfo {author} {\bibfnamefont {M.}~\bibnamefont {Scheibner}},\ and\ \bibinfo {author} {\bibfnamefont {C.-C.}\ \bibnamefont {Chien}},\ }\href@noop {} {\bibfield  {journal} {\bibinfo  {journal} {Europhysics Letters}\ }\textbf {\bibinfo {volume} {123}},\ \bibinfo {pages} {47002} (\bibinfo {year} {2018})}\BibitemShut {NoStop}%
\bibitem [{\citenamefont {Dugar}\ and\ \citenamefont {Chien}(2022)}]{dugar2022geometry}%
  \BibitemOpen
  \bibfield  {author} {\bibinfo {author} {\bibfnamefont {P.}~\bibnamefont {Dugar}}\ and\ \bibinfo {author} {\bibfnamefont {C.-C.}\ \bibnamefont {Chien}},\ }\href@noop {} {\bibfield  {journal} {\bibinfo  {journal} {Physical Review E}\ }\textbf {\bibinfo {volume} {105}},\ \bibinfo {pages} {064111} (\bibinfo {year} {2022})}\BibitemShut {NoStop}%
\bibitem [{\citenamefont {Aharonov}\ and\ \citenamefont {Bohm}(1959)}]{aharonov1959significance}%
  \BibitemOpen
  \bibfield  {author} {\bibinfo {author} {\bibfnamefont {Y.}~\bibnamefont {Aharonov}}\ and\ \bibinfo {author} {\bibfnamefont {D.}~\bibnamefont {Bohm}},\ }\href@noop {} {\bibfield  {journal} {\bibinfo  {journal} {Physical Review}\ }\textbf {\bibinfo {volume} {115}},\ \bibinfo {pages} {485} (\bibinfo {year} {1959})}\BibitemShut {NoStop}%
\bibitem [{\citenamefont {Roushan}\ \emph {et~al.}(2017)\citenamefont {Roushan}, \citenamefont {Neill}, \citenamefont {Megrant}, \citenamefont {Chen}, \citenamefont {Babbush}, \citenamefont {Barends}, \citenamefont {Campbell}, \citenamefont {Chen}, \citenamefont {Chiaro}, \citenamefont {Dunsworth} \emph {et~al.}}]{roushan2017chiral}%
  \BibitemOpen
  \bibfield  {author} {\bibinfo {author} {\bibfnamefont {P.}~\bibnamefont {Roushan}}, \bibinfo {author} {\bibfnamefont {C.}~\bibnamefont {Neill}}, \bibinfo {author} {\bibfnamefont {A.}~\bibnamefont {Megrant}}, \bibinfo {author} {\bibfnamefont {Y.}~\bibnamefont {Chen}}, \bibinfo {author} {\bibfnamefont {R.}~\bibnamefont {Babbush}}, \bibinfo {author} {\bibfnamefont {R.}~\bibnamefont {Barends}}, \bibinfo {author} {\bibfnamefont {B.}~\bibnamefont {Campbell}}, \bibinfo {author} {\bibfnamefont {Z.}~\bibnamefont {Chen}}, \bibinfo {author} {\bibfnamefont {B.}~\bibnamefont {Chiaro}}, \bibinfo {author} {\bibfnamefont {A.}~\bibnamefont {Dunsworth}}, \emph {et~al.},\ }\href@noop {} {\bibfield  {journal} {\bibinfo  {journal} {Nature Physics}\ }\textbf {\bibinfo {volume} {13}},\ \bibinfo {pages} {146} (\bibinfo {year} {2017})}\BibitemShut {NoStop}%
\bibitem [{\citenamefont {Fang}\ \emph {et~al.}(2012)\citenamefont {Fang}, \citenamefont {Yu},\ and\ \citenamefont {Fan}}]{fang2012realizing}%
  \BibitemOpen
  \bibfield  {author} {\bibinfo {author} {\bibfnamefont {K.}~\bibnamefont {Fang}}, \bibinfo {author} {\bibfnamefont {Z.}~\bibnamefont {Yu}},\ and\ \bibinfo {author} {\bibfnamefont {S.}~\bibnamefont {Fan}},\ }\href@noop {} {\bibfield  {journal} {\bibinfo  {journal} {Nature photonics}\ }\textbf {\bibinfo {volume} {6}},\ \bibinfo {pages} {782} (\bibinfo {year} {2012})}\BibitemShut {NoStop}%
\bibitem [{\citenamefont {Lin}\ \emph {et~al.}(2009)\citenamefont {Lin}, \citenamefont {Compton}, \citenamefont {Jim{\'e}nez-Garc{\'\i}a}, \citenamefont {Porto},\ and\ \citenamefont {Spielman}}]{lin2009synthetic}%
  \BibitemOpen
  \bibfield  {author} {\bibinfo {author} {\bibfnamefont {Y.-J.}\ \bibnamefont {Lin}}, \bibinfo {author} {\bibfnamefont {R.~L.}\ \bibnamefont {Compton}}, \bibinfo {author} {\bibfnamefont {K.}~\bibnamefont {Jim{\'e}nez-Garc{\'\i}a}}, \bibinfo {author} {\bibfnamefont {J.~V.}\ \bibnamefont {Porto}},\ and\ \bibinfo {author} {\bibfnamefont {I.~B.}\ \bibnamefont {Spielman}},\ }\href@noop {} {\bibfield  {journal} {\bibinfo  {journal} {Nature}\ }\textbf {\bibinfo {volume} {462}},\ \bibinfo {pages} {628} (\bibinfo {year} {2009})}\BibitemShut {NoStop}%
\bibitem [{\citenamefont {Kiefer}\ \emph {et~al.}(2019)\citenamefont {Kiefer}, \citenamefont {Hakelberg}, \citenamefont {Wittemer}, \citenamefont {Berm{\'u}dez}, \citenamefont {Porras}, \citenamefont {Warring},\ and\ \citenamefont {Schaetz}}]{kiefer2019floquet}%
  \BibitemOpen
  \bibfield  {author} {\bibinfo {author} {\bibfnamefont {P.}~\bibnamefont {Kiefer}}, \bibinfo {author} {\bibfnamefont {F.}~\bibnamefont {Hakelberg}}, \bibinfo {author} {\bibfnamefont {M.}~\bibnamefont {Wittemer}}, \bibinfo {author} {\bibfnamefont {A.}~\bibnamefont {Berm{\'u}dez}}, \bibinfo {author} {\bibfnamefont {D.}~\bibnamefont {Porras}}, \bibinfo {author} {\bibfnamefont {U.}~\bibnamefont {Warring}},\ and\ \bibinfo {author} {\bibfnamefont {T.}~\bibnamefont {Schaetz}},\ }\href@noop {} {\bibfield  {journal} {\bibinfo  {journal} {Physical Review Letters}\ }\textbf {\bibinfo {volume} {123}},\ \bibinfo {pages} {213605} (\bibinfo {year} {2019})}\BibitemShut {NoStop}%
\bibitem [{\citenamefont {Scully}\ and\ \citenamefont {Fleischhauer}(1992)}]{scully1992high}%
  \BibitemOpen
  \bibfield  {author} {\bibinfo {author} {\bibfnamefont {M.~O.}\ \bibnamefont {Scully}}\ and\ \bibinfo {author} {\bibfnamefont {M.}~\bibnamefont {Fleischhauer}},\ }\href@noop {} {\bibfield  {journal} {\bibinfo  {journal} {Physical review letters}\ }\textbf {\bibinfo {volume} {69}},\ \bibinfo {pages} {1360} (\bibinfo {year} {1992})}\BibitemShut {NoStop}%
\bibitem [{\citenamefont {Weymann}\ \emph {et~al.}(2011)\citenamefont {Weymann}, \citenamefont {Bu{\l}ka},\ and\ \citenamefont {Barna{\'s}}}]{weymann2011dark}%
  \BibitemOpen
  \bibfield  {author} {\bibinfo {author} {\bibfnamefont {I.}~\bibnamefont {Weymann}}, \bibinfo {author} {\bibfnamefont {B.}~\bibnamefont {Bu{\l}ka}},\ and\ \bibinfo {author} {\bibfnamefont {J.}~\bibnamefont {Barna{\'s}}},\ }\href@noop {} {\bibfield  {journal} {\bibinfo  {journal} {Physical Review B}\ }\textbf {\bibinfo {volume} {83}},\ \bibinfo {pages} {195302} (\bibinfo {year} {2011})}\BibitemShut {NoStop}%
\bibitem [{\citenamefont {Emary}(2007)}]{emary2007dark}%
  \BibitemOpen
  \bibfield  {author} {\bibinfo {author} {\bibfnamefont {C.}~\bibnamefont {Emary}},\ }\href@noop {} {\bibfield  {journal} {\bibinfo  {journal} {Physical Review B}\ }\textbf {\bibinfo {volume} {76}},\ \bibinfo {pages} {245319} (\bibinfo {year} {2007})}\BibitemShut {NoStop}%
\bibitem [{\citenamefont {Zhang}\ \emph {et~al.}(2018)\citenamefont {Zhang}, \citenamefont {Hu}, \citenamefont {Lin}, \citenamefont {Niu}, \citenamefont {Xia}, \citenamefont {Gong},\ and\ \citenamefont {Gong}}]{zhang2018thermal}%
  \BibitemOpen
  \bibfield  {author} {\bibinfo {author} {\bibfnamefont {S.}~\bibnamefont {Zhang}}, \bibinfo {author} {\bibfnamefont {Y.}~\bibnamefont {Hu}}, \bibinfo {author} {\bibfnamefont {G.}~\bibnamefont {Lin}}, \bibinfo {author} {\bibfnamefont {Y.}~\bibnamefont {Niu}}, \bibinfo {author} {\bibfnamefont {K.}~\bibnamefont {Xia}}, \bibinfo {author} {\bibfnamefont {J.}~\bibnamefont {Gong}},\ and\ \bibinfo {author} {\bibfnamefont {S.}~\bibnamefont {Gong}},\ }\href@noop {} {\bibfield  {journal} {\bibinfo  {journal} {Nature Photonics}\ }\textbf {\bibinfo {volume} {12}},\ \bibinfo {pages} {744} (\bibinfo {year} {2018})}\BibitemShut {NoStop}%
\bibitem [{\citenamefont {Scheucher}\ \emph {et~al.}(2016)\citenamefont {Scheucher}, \citenamefont {Hilico}, \citenamefont {Will}, \citenamefont {Volz},\ and\ \citenamefont {Rauschenbeutel}}]{scheucher2016quantum}%
  \BibitemOpen
  \bibfield  {author} {\bibinfo {author} {\bibfnamefont {M.}~\bibnamefont {Scheucher}}, \bibinfo {author} {\bibfnamefont {A.}~\bibnamefont {Hilico}}, \bibinfo {author} {\bibfnamefont {E.}~\bibnamefont {Will}}, \bibinfo {author} {\bibfnamefont {J.}~\bibnamefont {Volz}},\ and\ \bibinfo {author} {\bibfnamefont {A.}~\bibnamefont {Rauschenbeutel}},\ }\href@noop {} {\bibfield  {journal} {\bibinfo  {journal} {Science}\ }\textbf {\bibinfo {volume} {354}},\ \bibinfo {pages} {1577} (\bibinfo {year} {2016})}\BibitemShut {NoStop}%
\bibitem [{\citenamefont {Breuer}\ \emph {et~al.}(2002)\citenamefont {Breuer}, \citenamefont {Petruccione} \emph {et~al.}}]{breuer2002theory}%
  \BibitemOpen
  \bibfield  {author} {\bibinfo {author} {\bibfnamefont {H.-P.}\ \bibnamefont {Breuer}}, \bibinfo {author} {\bibfnamefont {F.}~\bibnamefont {Petruccione}}, \emph {et~al.},\ }\href@noop {} {\emph {\bibinfo {title} {The theory of open quantum systems}}}\ (\bibinfo  {publisher} {Oxford University Press on Demand},\ \bibinfo {year} {2002})\BibitemShut {NoStop}%
\bibitem [{\citenamefont {Landi}\ \emph {et~al.}(2022)\citenamefont {Landi}, \citenamefont {Poletti},\ and\ \citenamefont {Schaller}}]{landi2022nonequilibrium}%
  \BibitemOpen
  \bibfield  {author} {\bibinfo {author} {\bibfnamefont {G.~T.}\ \bibnamefont {Landi}}, \bibinfo {author} {\bibfnamefont {D.}~\bibnamefont {Poletti}},\ and\ \bibinfo {author} {\bibfnamefont {G.}~\bibnamefont {Schaller}},\ }\href@noop {} {\bibfield  {journal} {\bibinfo  {journal} {Reviews of Modern Physics}\ }\textbf {\bibinfo {volume} {94}},\ \bibinfo {pages} {045006} (\bibinfo {year} {2022})}\BibitemShut {NoStop}%
\bibitem [{\citenamefont {Deffner}\ and\ \citenamefont {Campbell}(2019)}]{deffner2019quantum}%
  \BibitemOpen
  \bibfield  {author} {\bibinfo {author} {\bibfnamefont {S.}~\bibnamefont {Deffner}}\ and\ \bibinfo {author} {\bibfnamefont {S.}~\bibnamefont {Campbell}},\ }\href@noop {} {\emph {\bibinfo {title} {Quantum Thermodynamics: An introduction to the thermodynamics of quantum information}}}\ (\bibinfo  {publisher} {Morgan \& Claypool Publishers},\ \bibinfo {year} {2019})\BibitemShut {NoStop}%
\bibitem [{\citenamefont {Plaszk{\'o}}\ \emph {et~al.}(2020)\citenamefont {Plaszk{\'o}}, \citenamefont {Rakyta}, \citenamefont {Cserti}, \citenamefont {Korm{\'a}nyos},\ and\ \citenamefont {Lambert}}]{plaszko2020quantum}%
  \BibitemOpen
  \bibfield  {author} {\bibinfo {author} {\bibfnamefont {N.~L.}\ \bibnamefont {Plaszk{\'o}}}, \bibinfo {author} {\bibfnamefont {P.}~\bibnamefont {Rakyta}}, \bibinfo {author} {\bibfnamefont {J.}~\bibnamefont {Cserti}}, \bibinfo {author} {\bibfnamefont {A.}~\bibnamefont {Korm{\'a}nyos}},\ and\ \bibinfo {author} {\bibfnamefont {C.~J.}\ \bibnamefont {Lambert}},\ }\href@noop {} {\bibfield  {journal} {\bibinfo  {journal} {Nanomaterials}\ }\textbf {\bibinfo {volume} {10}},\ \bibinfo {pages} {1033} (\bibinfo {year} {2020})}\BibitemShut {NoStop}%
\bibitem [{\citenamefont {Baselmans}\ \emph {et~al.}(1999)\citenamefont {Baselmans}, \citenamefont {Morpurgo}, \citenamefont {Van~Wees},\ and\ \citenamefont {Klapwijk}}]{baselmans1999reversing}%
  \BibitemOpen
  \bibfield  {author} {\bibinfo {author} {\bibfnamefont {J.}~\bibnamefont {Baselmans}}, \bibinfo {author} {\bibfnamefont {A.}~\bibnamefont {Morpurgo}}, \bibinfo {author} {\bibfnamefont {B.}~\bibnamefont {Van~Wees}},\ and\ \bibinfo {author} {\bibfnamefont {T.}~\bibnamefont {Klapwijk}},\ }\href@noop {} {\bibfield  {journal} {\bibinfo  {journal} {Nature}\ }\textbf {\bibinfo {volume} {397}},\ \bibinfo {pages} {43} (\bibinfo {year} {1999})}\BibitemShut {NoStop}%
\bibitem [{\citenamefont {Baselmans}\ \emph {et~al.}(2002)\citenamefont {Baselmans}, \citenamefont {Heikkil{\"a}}, \citenamefont {Van~Wees},\ and\ \citenamefont {Klapwijk}}]{baselmans2002direct}%
  \BibitemOpen
  \bibfield  {author} {\bibinfo {author} {\bibfnamefont {J.}~\bibnamefont {Baselmans}}, \bibinfo {author} {\bibfnamefont {T.}~\bibnamefont {Heikkil{\"a}}}, \bibinfo {author} {\bibfnamefont {B.}~\bibnamefont {Van~Wees}},\ and\ \bibinfo {author} {\bibfnamefont {T.}~\bibnamefont {Klapwijk}},\ }\href@noop {} {\bibfield  {journal} {\bibinfo  {journal} {Physical review letters}\ }\textbf {\bibinfo {volume} {89}},\ \bibinfo {pages} {207002} (\bibinfo {year} {2002})}\BibitemShut {NoStop}%
\bibitem [{\citenamefont {Hofer}\ \emph {et~al.}(2017)\citenamefont {Hofer}, \citenamefont {Perarnau-Llobet}, \citenamefont {Miranda}, \citenamefont {Haack}, \citenamefont {Silva}, \citenamefont {Brask},\ and\ \citenamefont {Brunner}}]{hofer2017markovian}%
  \BibitemOpen
  \bibfield  {author} {\bibinfo {author} {\bibfnamefont {P.~P.}\ \bibnamefont {Hofer}}, \bibinfo {author} {\bibfnamefont {M.}~\bibnamefont {Perarnau-Llobet}}, \bibinfo {author} {\bibfnamefont {L.~D.~M.}\ \bibnamefont {Miranda}}, \bibinfo {author} {\bibfnamefont {G.}~\bibnamefont {Haack}}, \bibinfo {author} {\bibfnamefont {R.}~\bibnamefont {Silva}}, \bibinfo {author} {\bibfnamefont {J.~B.}\ \bibnamefont {Brask}},\ and\ \bibinfo {author} {\bibfnamefont {N.}~\bibnamefont {Brunner}},\ }\href@noop {} {\bibfield  {journal} {\bibinfo  {journal} {New Journal of Physics}\ }\textbf {\bibinfo {volume} {19}},\ \bibinfo {pages} {123037} (\bibinfo {year} {2017})}\BibitemShut {NoStop}%
\bibitem [{\citenamefont {Marian}\ and\ \citenamefont {Marian}(2012)}]{marian2012uhlmann}%
  \BibitemOpen
  \bibfield  {author} {\bibinfo {author} {\bibfnamefont {P.}~\bibnamefont {Marian}}\ and\ \bibinfo {author} {\bibfnamefont {T.~A.}\ \bibnamefont {Marian}},\ }\href@noop {} {\bibfield  {journal} {\bibinfo  {journal} {Physical Review A}\ }\textbf {\bibinfo {volume} {86}},\ \bibinfo {pages} {022340} (\bibinfo {year} {2012})}\BibitemShut {NoStop}%
\end{thebibliography}
\providecommand{\noopsort}[1]{}\providecommand{\singleletter}[1]{#1}%

\begin{center}

\widetext
\newpage
\textbf{\large Supplemental Materials: Quantum Control of Heat Current}
\end{center}
\vspace{1cm}

\section{LQME}
\setcounter{equation}{1}
\renewcommand{\theequation}{A\thesection.\arabic{equation}}
Due to the linearized dynamics and zero mean Gaussian nature of the quantum noise, the system retains its Gausianity and thus can be fully expressed by a $2n\times2n$ ($n=3)$ covariance matrix, whose elements are defined as
\begin{equation}
V_{lm}(t) = \langle u_l(t) u_m(t) + u_m(t) u_l(t) \rangle .
\end{equation}
Here $u_i(t)$ is an element of the $2n$ dimensional vector $u(t)=[x_2(t),x_1(t),x_1(t),p_2(t),p_1(t),p_3(t)]^T$, with $x_i=(a_i+a^\dagger_i)/2$, $p_i=(a_i-a^\dagger_i)/2i$ and $[x_i,p_i]=i$

In the steady state, we can solve the LQME using the following equation,
\begin{equation}
    R V(t=\infty) + V (t=\infty) R^T =  D,
\end{equation}

where \begin{align}\label{R}
 R=\left(
 \begin{array}{cccccc}
 \gamma_2/2 & iJ_{21} & iJ_{23} & i\gamma_2/2 & -J_{21} & -J_{23}\\
 iJ_{12} & \gamma_1/2 & iJ_{13} e^{i\theta} & -J_{12} & i\gamma_1/2 & -J_{13}e^{i\theta}\\
 iJ_{32}  & iJ_{31} e^{-i\theta} & \gamma_3/2 & -J_{32} & -J_{31}e^{-i\theta} & i\gamma_3/2 \\
 -i\gamma_2/2  & J_{21} & J_{23} &\gamma_2/2 &  iJ_{21} & iJ_{23} \\
 J_{12} & -i\gamma_1/2 & J_{13} e^{i\theta} & iJ_{12} & \gamma_1/2 & iJ_{13}e^{i\theta}\\
  J_{32}  & J_{31} e^{-i\theta} & -i\gamma_3/2 & iJ_{32} & iJ_{31}e^{-i\theta} & \gamma_3/2 \\
 \end{array}
 \right),
\end{align}
and \begin{align}\label{D}
 D=\left(
 \begin{array}{cccccc}
 \gamma_2 N_{H} & 0 & 0 & 0 & 0 & 0\\
 0 & \gamma_1 N_{C} & 0  & 0 &0 & 0\\
 0  & 0 & \gamma_3 N_{C} & 0 & 0 & 0 \\
 0  & 0 & 0 &\gamma_2 N_{H} &  0 & 0 \\
 0 & 0 & 0 & 0 & \gamma_1 N_{C} & 0\\
  0  & 0 & 0 & 0 & 0 & \gamma_3 N_{C} \\
 \end{array}
 \right),
\end{align}

\section{Exact Numerics}
\setcounter{equation}{0}
\renewcommand{\theequation}{B\thesection.\arabic{equation}}

In this section, we present our methodology for obtaining the exact numerics. The total Hamiltonian (including system and the baths) can be written as,
\begin{align}\label{Ham_tot}
        &\emph{$\tilde{H}$} = \emph{H} +  \sum_{\alpha=c,h} \left[ \emph{H}_{\alpha}+\emph{V}_{\alpha}\right] \\
        &\emph{H} = \sum_{l=1} ^3 \omega a_l ^\dagger a_l +  \sum_{lm} J_{lm} \left(e^{i \theta_{lm}} a_l ^ \dagger a_m + e^{-i \theta_{lm}} a_l a_m ^\dagger \right),\nonumber\\  
        &\emph{H}_{\alpha} = \sum_{j}\omega_{\alpha, j}b_{\alpha, j}^\dagger b_{\alpha, j},\nonumber\\
        &\emph{V}_{h} = \sum_{j}g_{h, j}\left(b_{h, j}^\dagger a_2 + a_2^\dagger b_{h, j}\right),\nonumber\\
        &\emph{V}_{c} = \sum_{j}g_{c, j}\left(b_{c, j}^\dagger (a_1 + a_3) + (a_1^\dagger +  a_3^\dagger) b_{c, j}\right)\nonumber,  
\end{align}
where $b_{\alpha, j}$ $b_{\alpha, j}^\dagger$ denote annihilation (creation) operators of the baths, $\omega_{\alpha, j}$ are the frequencies corresponding to the bath modes, and $g_{\alpha, j}$ denote the interaction strength between system and bath modes. The Hamiltonian in Eq.\ref{Ham_total}, describes a Gaussian system and our analysis revolves around the evolution of the covariance matrix. Our focus is on the unitary evolution of the whole compound, i.e. system with finite yet large number of oscillators in the baths. To this end, we delve into the time evolution of all system and bath modes and we write them compactly as,
\begin{align*}
    A = (a_2, b_{h, 1}, b_{h,2}, \ldots, b_{h,N}, a_1, a_3, b_{c, 1}, b_{c,2}, \ldots, b_{c,N})^T
\end{align*} and 
\begin{align*}
    A^{\dagger} = (a_2^{\dagger}, b_{h, 1}^{\dagger}, b_{h,2}^{\dagger}, \ldots, b_{h,N}^{\dagger}, a_1^{\dagger}, a_3^{\dagger}, b_{c, 1}^{\dagger}, b_{c,2}^{\dagger}, \ldots, b_{c,N}^{\dagger})^T,
\end{align*} where N represents the number of bath oscillators. The time evolution of the modes is given by,
\begin{align}
&A(t) = e^{-iWt}A(0),\\ 
&A^\dagger(t) = e^{i\overline{W}t}A^\dagger(0).
\end{align}
$\overline{W}$ is the conjugate of the complex matrix $W$ given by,
\begin{equation}
W=
\begin{bmatrix}
\omega  &  g_{h, 1} & g_{h, 2} &\cdots & g_{h, N} & J & J & 0 & 0  & \cdots & 0 \\
g_{h, 1}  & \omega_{h, 1} & 0 & \cdots &\cdots &\cdots&\cdots  &&  &\cdots& 0\\
g_{h, 2} & 0  & \omega_{h, 2}& 0 &  \cdots &\cdots&\cdots  &&  & \cdots& 0 \\
\vdots & 0 & 0 & \ddots & 0 &\cdots &\cdots  &&  &\cdots&  0 \\
g_{h, N}  & 0& \cdots & \cdots & \omega_{h, N}  & 0   & \cdots &  &   &\cdots& 0\\
J  & 0 & \cdots& &  0 & \omega  & Je^{i\theta}  & g_{c, 1} & g_{c, 2} & \cdots & g_{c, N}\\
J  & 0 & \cdots& &  0 & Je^{-i\theta} & \omega    & g_{c, 1} & g_{c, 2} & \cdots & g_{c, N}\\
0  & 0 & \cdots& &  0 & g_{c, 1} & g_{c, 1}    & \omega_{c, 1} & 0 & \cdots & 0\\
\vdots  & \vdots & \cdots & &  0 & g_{c, 2} & g_{c, 2}& 0 &\omega_{c, 2}  & \cdots & 0\\
\vdots  & \vdots & \cdots & &  0 &\vdots & \vdots& 0 &0  & \ddots & 0\\
0  & 0 & \cdots & &  0 &g_{c, N} & g_{c, N}& 0 &0  & \cdots & \omega_{c, N}\\
\end{bmatrix}
\end{equation}
The time evolution of the position $X = \frac{1}{2}\left(A+A^{\dagger}\right)$ and momentum $P = \frac{1}{2i}\left(A-A^{\dagger}\right)$ can be given by,
\begin{align}
    X(t) &= \psi_R X(0) - \psi_I P(0)\\ \nonumber
    P(t) &= \psi_I X(0) + \psi_R P(0)
\end{align}
where $\psi_R$ and $\psi_I$ are given by,
\begin{align}
    \psi_R &= (e^{-iWt} + e^{i\overline{W}t})/2 \\ \nonumber
    \psi_I &= (e^{-iWt} - e^{i\overline{W}t})/2i \\
\end{align}
The time evolution of the vector, 
\begin{equation*}
    U = \left( x_2, x_{h,1}, x_{h,2},\ldots, x_{h,N}, x_1, x_3, x_{c,1}, x_{c,2},\ldots,x_{c,N},\\ p_2, p_{h,1}, p_{h,2},\ldots, p_{h,N}, p_1, p_3, p_{c,1}, p_{c,2},\ldots, p_{c,N}\right)^T
\end{equation*} is given by,
\begin{equation}
    U(t) = \mathcal{M} R(0)
\end{equation}
where where $x_j, p_j$ are the position and momentum of the $j^{th}$ oscillator and \\ $\mathcal{M}=\begin{pmatrix}
\psi_R & -\psi_I \\
\psi_I & \psi_R 
\end{pmatrix}$.

Since the system is Gaussian we are interested in the time evolution of the covariance matrix
\begin{align}
    \mathcal{C}_{i,j}(t) &= \langle U_i(t) U_j(t) + U_j(t) U_i(t) \rangle \\ \nonumber
    &= \sum_{q,r} \mathcal{M}_{i,q} \mathcal{M}_{j,r} \mathcal{C}_{q,r}(0)
\end{align}
The evolution of expectation values of pairwise correlations can be obtained from the elements of the evolved covariance matrix as follows,\\
\begin{align}
    \Re(\langle a_1^{\dagger}(t)a_3(t) \rangle) = \Re(\langle a_1^{\dagger}(t)a_3(t) \rangle) = & \mathcal{C}_{N+2,N+3}(t)\\ \nonumber
    \Im(\langle a_1^{\dagger}(t)a_3(t) \rangle) = -\Im (\langle a_3^{\dagger}(t)a_1(t) \rangle) = & \mathcal{C}_{N+2,3N+6}(t)\\ \nonumber
    \Re(\langle a_2^{\dagger}(t)a_1(t) \rangle) = \Re(\langle a_1^{\dagger}(t)a_2(t) \rangle) = & \mathcal{C}_{1,N+2}(t)\\ \nonumber
    \Im(\langle a_2^{\dagger}(t)a_1(t) \rangle) = -\Im( \langle a_1^{\dagger}(t)a_2(t) \rangle) = & \mathcal{C}_{1,3N+5}(t)\\ \nonumber
    \Re(\langle a_2^{\dagger}(t)a_3(t) \rangle) = \Re(\langle a_3^{\dagger}(t)a_2(t) \rangle) = & \mathcal{C}_{1,N+3}(t)\\ \nonumber
    \Im(\langle a_2^{\dagger}(t)a_3(t) \rangle) = -\Im( \langle a_3^{\dagger}(t)a_2(t) \rangle) = & \mathcal{C}_{1,3N+6}(t)\\ \nonumber
\end{align}
\subsection{Parameters }
 In our simulation, we consider a bath consisting of $N$ oscillators ($j=1,2,...,N$), with the bath modes $\omega_{\alpha, j}$ evenly spaced over a range from $\omega_c/N$ to $\omega_c$, where $\omega_c$ represents the cut-off frequency of the ohmic spectral density. Specifically, the bath modes are defined as:

\begin{equation}
\omega_{\alpha,j} = \frac{j}{N}\omega_c
\end{equation}

Here, $\alpha=H,C$, and $j=1,2,\ldots,N$.

The coupling strengths $g_{\alpha,j}$ capture the interaction between the system and the individual oscillators in the bath. For an ohmic distribution of bath oscillators, we employ the following spectral density function,

\begin{equation}\label{ohmic}
J\left(\Omega\right) = \kappa \Omega e^{-\Omega/\omega_c} = \sum_{j=1}^N g_{\alpha,j}^2\delta(\Omega-\omega_{\alpha,j})
\end{equation}

where $\kappa$ is a constant governing the interaction strengths. In the low-frequency regime ($\Omega\leq\omega_c$), we can approximate this spectral density as

\begin{equation}\label{ohmic_appx}
J\left(\Omega\right) \approx \kappa \Omega
\end{equation}

By integrating Eq. \ref{ohmic} over the discrete frequency domain, we relate the coupling strengths to the spectral density,

\begin{equation}\label{ohmic_to_strngth}
\sum_{j=1}^NJ\left(\omega_{\alpha,j}\right)\frac{\omega_c}{N}
=\sum_{j=1}^N g_{\alpha,j}^2
\end{equation}

Utilizing the low-frequency approximation in Eq. \ref{ohmic_appx} and Eq. \ref{ohmic_to_strngth}, an expression for the coupling strengths can be given by,

\begin{equation}\label{gk}
g_{\alpha,j}^2 \approx j \kappa \left( \frac{\omega_c}{N} \right)^2
\end{equation}

The system-bath coupling rate $\gamma$ is directly proportional to the spectral density of the baths, given by $\gamma=2\pi J\left(\omega\right)\approx2\pi\kappa\omega$. Substituting this expression into Eq. \ref{gk}, we rewrite the coupling strengths as,

\begin{equation}
g_{\alpha,j} \approx \sqrt{j \frac{\gamma}{2\pi\omega}}\frac{\omega_c}{N}
\end{equation}
The initial state for the exact simulation is taken as, $\rho_{0} = \rho_{h}\otimes\rho_S\otimes\rho_{c}$, where $\rho_\alpha =\mathrm{exp}\left(-\frac{\emph{H}_\alpha }{T_\alpha} \right)/Tr\left[\mathrm{exp}\left(-\frac{\emph{H}_\alpha }{T_\alpha} \right)\right]$ and  $\rho_S=\otimes \rho_l$ with $\rho_l =\mathrm{exp}\left(-\frac{\omega a_l ^\dagger a_l }{T_l} \right)/Tr\left[\mathrm{exp}\left(-\frac{\omega a_l ^\dagger a_l }{T_l}\right)\right]$, $T_2 = T_h$ and $T_1=T_3=T_c$.

\subsection{Benchmark}
To ascertain the system's steady state via exact simulation, it becomes imperative to find a critical time ($\tau_{ss}$) where the oscillation of the system modes ceases reaching a quasi-steady state like behaviour before oscillation recurs. Importantly, $\tau_{ss}$ falls within the temporal bounds of the equilibration time $\tau_{eq}$ and the recurrence time $\tau_{rec}$. $\tau_{eq} = \frac {1} {\gamma} $ characterizes the timeframe for the system to equilibrate. The recurrence time $\tau_{rec}$ follows a linear scaling with the number of bath oscillators. Consequently, by having a large oscillator count in the bath (N=400), we ensure $\tau_{rec} \gg \tau_{eq}$. In our simulation, $\tau_{ss}\approx6 \tau_{eq}$, which is robust to all $J/\gamma$ values.

\begin{figure}[!t]
   \begin{center}
   \includegraphics[width=16 cm]{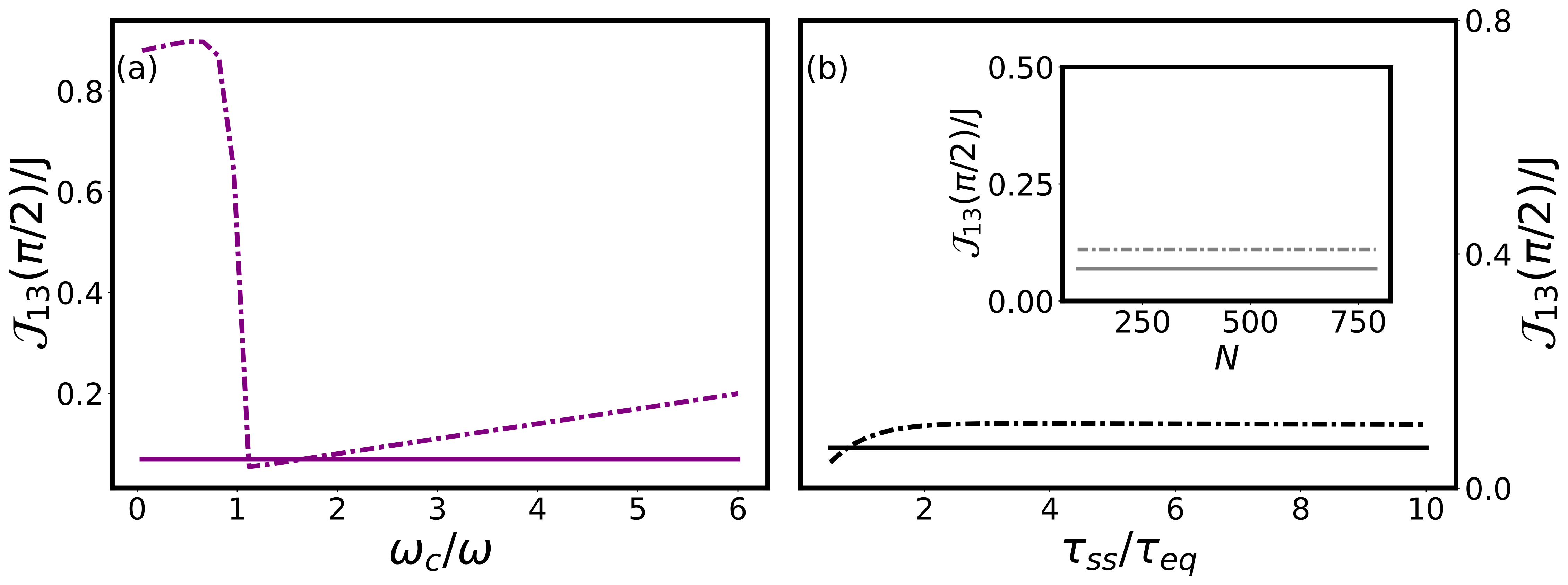}
   \caption{Benchmark for exact simulation: $\mathcal{J}_{13}\left(\pi/2\right)$/J from LQME (solid) and exact simulation (dash-dotted) as function of (a) $\omega_c/\omega$, (b) $\tau_ss/\tau_{eq}$ and N (inset) for $J/\gamma=0.1$.}.
   \label{fig:SM1}    
   \end{center}
  \end{figure}

We employed exact numeric simulations to benchmark our findings at low $J/\gamma$ values from the Lindblad quantum master equation and to extend the study to the large coupling regime. For the simulation, the cut-off frequency of the ohmic spectral density of the baths, the time for the evolution of all the modes, and the number of oscillators in the baths are chosen as $\omega_c = 3\omega$, $\tau_{ss}=6\tau_{eq}$, and $N=400$, respectively. These parameters were selected following an extensive benchmark analysis of the current $\mathcal{J}_{13}(\theta=\pi/2)$ from both the Lindblad master equation and exact numerical simulation as shown in Fig.~\ref{fig:SM1}. Internal current $\mathcal{J}_{13}(\theta=\pi/2)$ is plotted against $\omega_c/\omega$, $\tau_{ss}/\tau_{eq}$, and $N$ for a small coupling strength of $J/\gamma=0.1$. This benchmarking is crucial, particularly for such low coupling strengths, where we expect both the Lindblad master equation and exact numerical calculations to show high fidelity. Interestingly, for $\omega_c > 1.5\omega$, good agreement between the Lindblad master equation and exact numerical calculations is found, with minor discrepancies arising as $\omega_c$ increases. To capture all possible enrgy transitions in the system (harmonic oscillators 1,2 and 3) we select $\omega_c = 3$. For $\tau_{ss}>2\tau_{eq}$, the results show favourable agreement, and the result is independent of the number of oscillators in the bath ($N$) within the range of 100 to 800. This also emphasizes that $\tau_{ss}\approx6\tau_{eq}$ falls well within the range of ($\tau_{eq}$, $\tau_{rec}$).

\section{Fidelity}

In this section, we measure the fidelity between the states of the subsystem 1 and 3 evolved by exact numerics and LQME. Although the goal of our work is not a comaprative study between exact numerics and LQME for our system, we employ such measures mainly to understand any disparity between the above given descriptions un presence of complex interactions in the system.

For any two states $\rho_1$ and $\rho_2$ the fidelity is defined as,
\begin{equation}
    \mathcal{F}\left(\rho_1,\rho_2 \right) = \text{Tr}\left(\sqrt{\sqrt{\rho_1}\rho_2\sqrt{\rho_1}} \right)
\end{equation}

For two mode gaussian states given by $\mathcal{C}_1$ and $\mathcal{C}_2$, the fidelity is given by.
\begin{equation}
    \mathcal{F}\left(\mathcal{C}_1,\mathcal{C}_2\right)=\frac{1}{\sqrt{\Lambda}+\sqrt{\Phi}-\sqrt{\left(\sqrt{\Lambda}+\sqrt{\Phi}\right)^2-\Omega}}
\end{equation}
where,

$\Omega= \text{det}(\mathcal{C}_1 + \mathcal{C}_2)$, $\Lambda=2^4\text{det}(T\mathcal{C}_1\mathcal{C}_2-\mathbb{I}/4)$ and $\Phi=2^4\text{det}(\mathcal{C}_1+iT/2)\text{det}(\mathcal{C}_2+iT/2)$. Here $\mathbb{I}$ is the $4\times4$ identity matrix and the $T$ is given by $T_{lm}=-i\langle[u_l,u_m]\rangle$.

\section{Systematic error}
The discussion till now has been restricted to an idealistic situation, with a symmetric choice of system-bath parameters. However, in practice, there could be systematic errors ($\delta$, $\epsilon$), incorporated through 
\begin{align}
    J_{21} &= J(1+\delta), \hspace{0.2cm} J_{23} = J(1-\delta),\\
    \gamma_{1} & =\gamma(1+\epsilon), \hspace{0.45cm} \gamma_{3} = \gamma(1-\epsilon).
\end{align}

\begin{figure}[!t]
\begin{center}
\includegraphics[width = 16 cm]{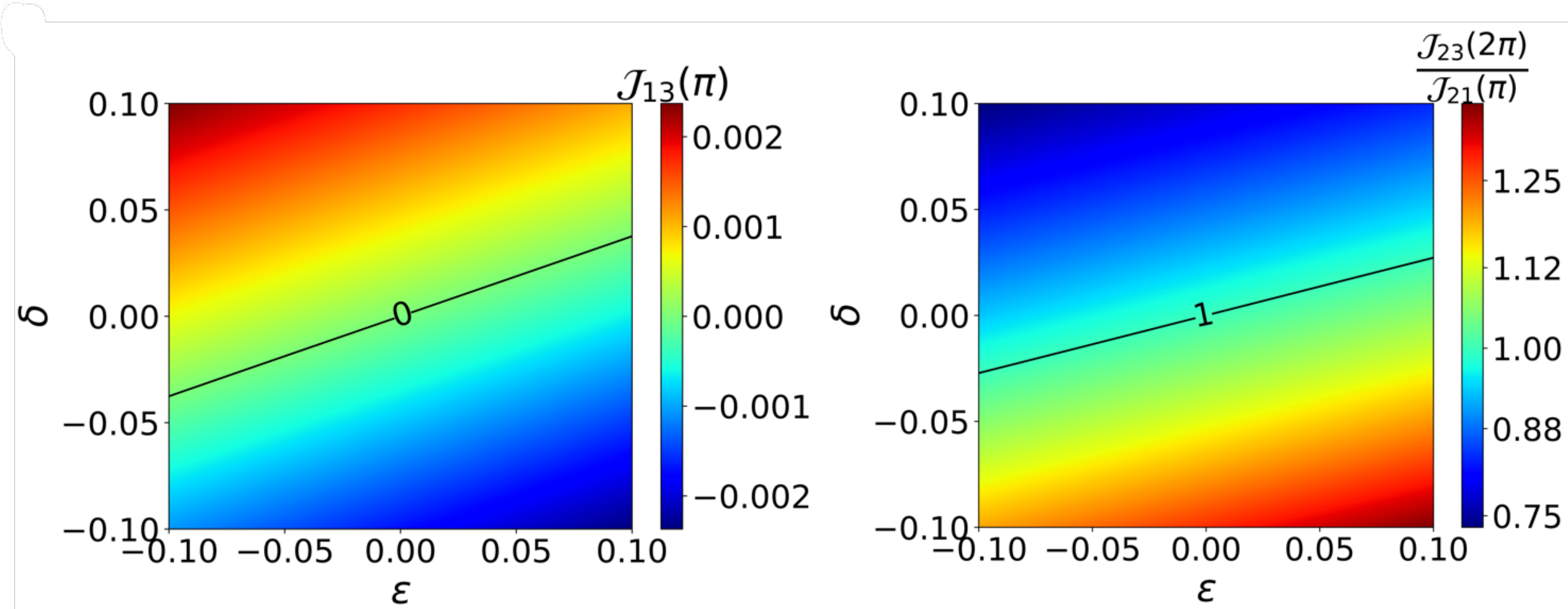}
\caption{Systematic error: (a) Color-coded plot for the values of $\mathcal{J}_{13}\left(\theta\right)$ at $\theta=\pi$ for different values of systematic error with relative strengths of ($\delta$, $\epsilon$). (b) Color-coded plot for the values of $\mathcal{J}_{23}\left(\theta+\pi\right)/\mathcal{J}_{21}\left(\theta\right)$ at $\theta=\pi$ for different values of systematic error with relative strengths of ($\delta$, $\epsilon$).}
\label{fig:SM2}    
\end{center}
\end{figure}

To demonstrate the impact of such errors on the switch and swap functionalities, we plot $\mathcal{J}_{13}(\pi)$ and $\mathcal{J}_{23}(2\pi)/\mathcal{J}_{21}(\pi)$ in the ($\epsilon$, $\delta$) plane. It is to be noted that due to the incoherent coupling arising through $\epsilon$ we observe a finite $\mathcal{J}_{13}$ at $\theta=\pi$. However, from Fig.~\ref{fig:SM2}, we find a range of ($\delta$, $\epsilon$) for which the functionalities remain robust against the errors. (shown as the black line in the plot).

\end{document}